\documentclass[conference]{IEEEtran}
\IEEEoverridecommandlockouts

\usepackage{cite}
\usepackage{amsmath,amssymb,amsfonts}
\usepackage{algorithmic}
\usepackage{graphicx}
\usepackage{textcomp}
\usepackage{xcolor}
\usepackage[english]{babel}
\usepackage[backref]{hyperref}
\usepackage{fancyhdr}
\usepackage{subcaption}
\usepackage{tabularx}
\usepackage{array}
\usepackage{url}

\usepackage{float}

\usepackage[export]{adjustbox}

\usepackage[ruled,linesnumbered,commentsnumbered]{algorithm2e}

\def\BibTeX{{\rm B\kern-.05em{\sc i\kern-.025em b}\kern-.08em
    T\kern-.1667em\lower.7ex\hbox{E}\kern-.125emX}}

\begin{document}

\pagenumbering{arabic}

\title{Fire-Flyer AI-HPC: A Cost-Effective Software-Hardware Co-Design for Deep Learning}

\author{
\thanks{
    \vspace{-2em}
    \hspace{-\marginparsep}\hspace{0.3em}
        \begin{minipage}{0.5\textwidth}
            \rule{0.97\linewidth}{0.4pt}
            Authors are listed in alphabetical order of their surnames.
        \end{minipage}
}
\IEEEauthorblockN{Wei An, Xiao Bi, Guanting Chen, Shanhuang Chen,Chengqi Deng, Honghui Ding, Kai Dong, Qiushi Du,}

\IEEEauthorblockN{Wenjun Gao, Kang Guan, Jianzhong Guo, Yongqiang Guo, Zhe Fu, Ying He, Panpan Huang, Jiashi Li,}

\IEEEauthorblockN{Wenfeng Liang, Xiaodong Liu, Xin Liu, Yiyuan Liu, Yuxuan Liu, Shanghao Lu, Xuan Lu, Xiaotao Nie, }

\IEEEauthorblockN{Tian Pei, Junjie Qiu, Hui Qu, Zehui Ren, Zhangli Sha, Xuecheng Su, Xiaowen Sun, Yixuan Tan, Minghui Tang,}

\IEEEauthorblockN{Shiyu Wang, Yaohui Wang, Yongji Wang, Ziwei Xie, Yiliang Xiong, Yanhong Xu, Shengfeng Ye, Shuiping Yu,}

\IEEEauthorblockN{Yukun Zha, Liyue Zhang*, Haowei Zhang, Mingchuan Zhang, Wentao Zhang, Yichao Zhang, Chenggang Zhao,}

\IEEEauthorblockN{ Yao Zhao, Shangyan Zhou, Shunfeng Zhou, Yuheng Zou}
\\
\IEEEauthorblockA{ DeepSeek-AI, Beijing, China}
\href{mailto:research@deepseek.com}{research@deepseek.com}

}

\maketitle
\thispagestyle{fancy}
\lhead{}
\rhead{}
\chead{}
\lfoot{\footnotesize{
SC24, November 17-22, 2024, Atlanta, Georgia, USA
\newline 979-8-3503-5291-7/24/\$31.00 \copyright 2024 IEEE}}
\rfoot{}
\cfoot{}
\renewcommand{\headrulewidth}{0pt}
\renewcommand{\footrulewidth}{0pt}

\begin{abstract}
The rapid progress in Deep Learning (DL) and Large Language Models (LLMs) has exponentially increased demands of computational power and bandwidth. This, combined with the high costs of faster computing chips and interconnects, has significantly inflated  High Performance Computing (HPC) construction costs. To address these challenges, we introduce the Fire-Flyer AI-HPC architecture, a synergistic hardware-software co-design framework and its best practices. For DL training, we deployed the Fire-Flyer 2 with 10,000 PCIe A100 GPUs, achieved performance approximating the DGX-A100 while reducing costs by half and energy consumption by 40\%. We specifically engineered HFReduce to accelerate allreduce communication and implemented numerous measures to keep our Computation-Storage Integrated Network congestion-free. Through our software stack, including HaiScale, 3FS, and HAI-Platform, we achieved substantial scalability by overlapping computation and communication. Our system-oriented experience from DL training provides valuable insights to drive future advancements in AI-HPC.
\end{abstract}


\IEEEpubid{\makebox[\columnwidth]{\hfill} \hspace{\columnsep}\makebox[\columnwidth]{\hfill}}
\IEEEpubidadjcol

\section{Introduction}
In recent years, Deep Learning (DL)\cite{lecun2015deep} has developed rapidly and is widely used in image recognition, speech recognition, content generation, autonomous driving, and other areas. The rapid development of DL is fundamentally tied to the support rendered by data. Training with copious amounts of data demands massive computational resources. Relying on Moore's Law\cite{schaller1997moore}, computer speeds double every two years on average, but the pace of DL far exceeds this speed. In particular, Large Language Models (LLMs)\cite{brown2020language,Anthropic_Claude,Google_Bard,OpenAI_Chatgpt,achiam2023gpt} that have become popular in recent years have exploded the demand for computational resources and memory. The parameters of LLMs can reach tens to thousands of billions, requiring hundreds or thousands of GPUs for training. Although LLM training is challenging, the emerging capabilities resulting from more parameters have shown the benefits of continued model expansion. Since then, researchers have gone down the path of making models bigger and never looked back. To acquire more computational resources, people have to expand more nodes. This leads to a surge in the cost of building AI infrastructure. How to reduce the cost of new data centers and how to build cost-effective clusters are also hot and challenging problems. Moreover, more nodes lead to higher energy consumption, which contradicts this era's goal of reducing carbon emissions and achieving carbon neutrality. Reducing energy consumption is also a challenging problem.

In this paper,  leverage our practical experience accumulated over the years to propose cost-effective strategies cost-effective strategies for constructing AI-HPC systems suitable for deep learning and LLMs.

\subsubsection* {\textbf{Fire-Flyer AI-HPC Architecture}} 
We have deployed a cluster composed of 10,000 PCIe A100 GPUs for Deep Learning training purposes. Details about the GPU nodes and network topology are provided in Section \ref{fireflyer2}, where we compare our architecture to the NVIDIA DGX-A100 \cite{nvidia_dgx}  in terms of cost-effectiveness and lower CO\textsubscript{2} emissions. In contrast, we must invest more in software optimizations to address the performance challenges of the PCIe architecture. The following sections will discuss about software-hardware co-design.
\subsection* {Key Technical Topics in our Architecture}
\begin{itemize}
\item \textbf{Network Co-Design}: The Two-Layer Fat-Tree Network\cite{fat_tree_6312192} integrates storage and computation network, as shown in Section \ref{network_topo_section}. The entire network is divided into two zones, and the platform supports cross-zone tasks. To prevent congestion, we employed various network tunings detailed in Section  \ref{network_tuning_section}.
\item \textbf{HFReduce}: Achieves computation-communication overlap via asynchronous allreduce on the CPU, outperforming NVIDIA Collective Communications Library (NCCL)\cite{NCCL_LINK} on our PCIe architecture, as discussed in Section \ref{HFReduce_section}.
\item \textbf{HaiScale}: As described in Section \ref{HaiScale_section}, optimizes parallelism methods for our PCIe architecture, such as Data Parallelism (DP), Pipeline Parallelism (PP)\cite{DBLP:journals/corr/abs-1811-06965}\cite{10.1145/3341301.3359646}, Tensor Parallelism (TP)\cite{shoeybi_megatron-lm_2020}\cite{10.1145/3458817.3476209}, Experts Parallelism (EP) \cite{10.1145/3577193.3593704,ae449111733a42c5980594f9133812c8,hwang2023tutel}, Fully Sharded Data Parallel(FSDP)\cite{zhao2023pytorch} and Zero Redundancy Optimizer(ZeRO)\cite{rajbhandari2020zero}.
\item \textbf{3FS Distributed File System}: Addresses I/O bottlenecks in big data AI tasks, configures with our communication and network tuning, reduces congestion in storage and computation integrated network topology,  as detailed in Section \ref{3fs_section}.
\item \textbf{HAI Platform}\cite{HAI_PLATFORM}: Offers task scheduling, fault handling, and disaster recovery, enhancing utilization and reducing costs. It provides an open-box solution for deep-learning researchers. It is already open-sourced: \href{https://github.com/HFAiLab/hai-platform}{https://github.com/HFAiLab/hai-platform}
\end{itemize}

\subsubsection* {\textbf{Stability and Robustness}} These are crucial topics in HPC. Our systems are equipped with robust mechanisms to handle hardware failures, minimizing downtime and impact on operations. These mechanisms, discussed in Section \ref{stability_section}, include:
\begin{itemize}
\item Disaster recovery through our checkpoint manager
\item A validator utility for detecting hardware failures
\item An overview of real hardware failure data from our cluster over the past year.
\end{itemize}
We hope these insights will be beneficial to industry peers and researchers alike.

\subsubsection* {\textbf{Discussion and Future Work}} In Section \ref{discuss_section}, we address some common questions regarding PCIe architecture, such as congestion control, maintenance cost, and stability compared with other architectures. In Section \ref{nextgen_section}, we propose the next generation of PCIe architecture, which is aimed at Mixture-of-Experts Large Language Models training and primarily utilizes multi-NICs and a Multi-Plane network.

\section{Background}
\subsection{Evolution of Deep Learning}
The revolution in Machine Learning and Deep Learning began in 2012 with AlexNet\cite{krizhevsky2012imagenet}, which outperformed traditional methods in image classification, marking the onset of big data utilization and increased computational demands. The emergence of ResNet\cite{he2016deep}, with its deeper layers, further broadened the horizons of image processing, truly bringing about the ``deep" in ``Deep Learning". At the same time, big data-driven model training nudged the evolution of data storage technology, leading to the advent of all-flash SSD distributed file systems.

Fast forward to 2017, Google's Transformer\cite{vaswani2017attention} made its grand entry, introducing the concept of "Attention is all you need", shaking up the field of Natural Language Processing (NLP). With the advent of more complex models like AlphaFold\cite{senior2020improved} and AlphaZero\cite{silver2018general} highlighting the need for more computational power and memory, revealing the limitations of traditional FP64 / FP32 computing devices.

Entering the 2020s saw the rise of LLMs as a game-changer in the AI sector. Research indicates that an upscale in the number of language model parameters and computational budget can significantly enhance model performance, given adequate training data. Consequently, despite requiring colossal computational resources, efforts are being made to train large models on tens or hundreds of billions or even trillions of parameters. Pioneering examples include GPT-3\cite{floridi2020gpt} and PaLM\cite{chowdhery2023palm}, which occupy close to 1TB of GPU Memory. Recognizing the potential, industry giants have set up large AI clusters to train LLMs while constantly investing in computational power chips.

The shift towards the Mixture-of-Experts (MoE) Models\cite{jacobs1991adaptive,jordan1994hierarchical,shazeer2017outrageously} architecture starting from GPT-4\cite{achiam2023gpt}, and the recent AI Generated Content (AIGC) multi-modal (Sora\cite{videoworldsimulators2024}) has amplified the demand for memory and computational resources. However, as AI development outpaces hardware development, leading to skyrocketing training costs, adopting cost-saving solutions has become imperative.

\begin{figure}[bt]
\centerline{\includegraphics[width=0.5\textwidth]{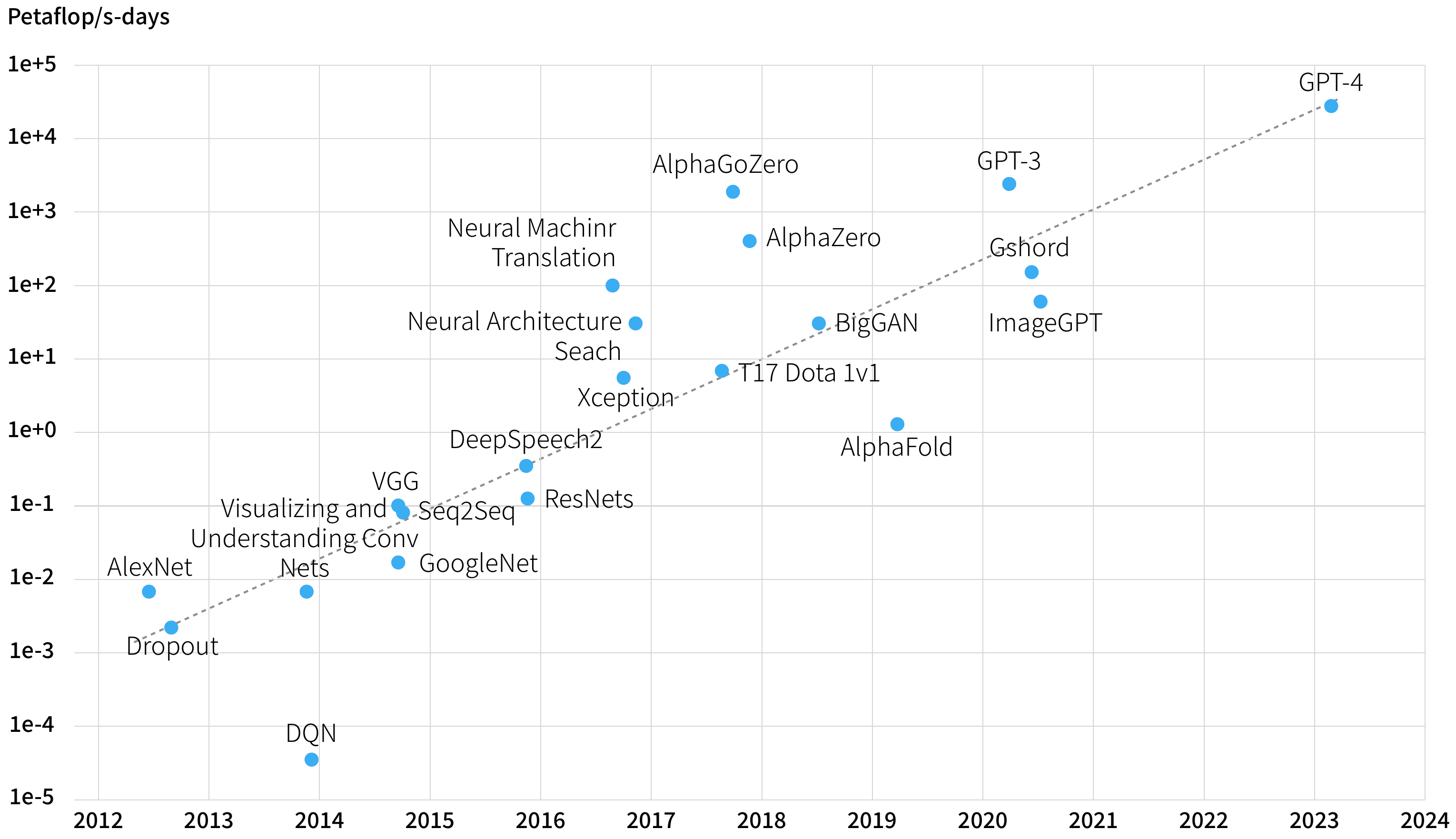}}
\caption{The Exponential Growth of Computational Power for Deep Learning.}
\label{growth_of_dl}
\vspace{-15pt} 
\end{figure}

\begin{figure}
    \centering
    \includegraphics[width=1\linewidth]{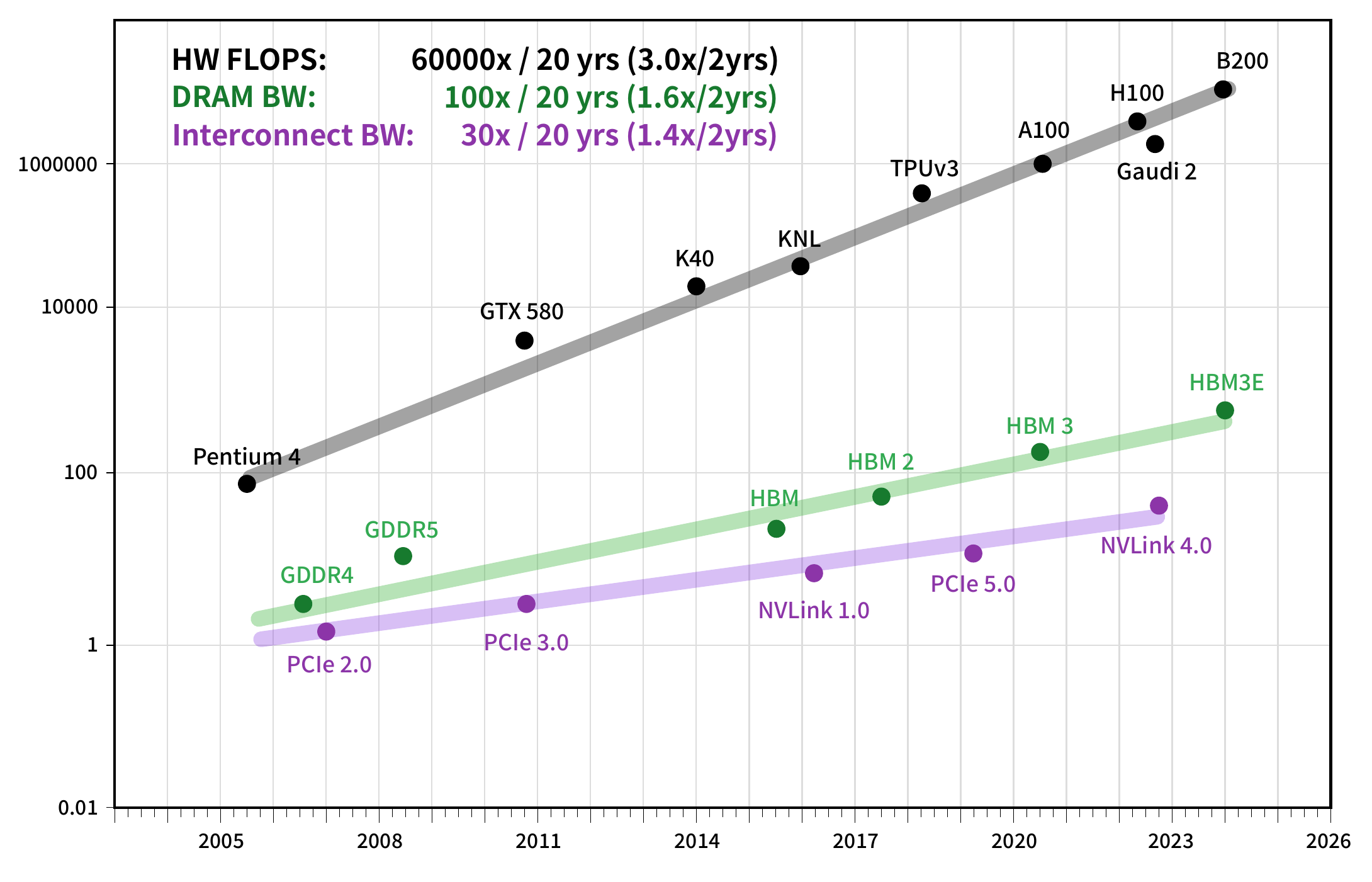}
\caption{Scaling of Peak Hardware FLOPS, and Memory/Interconnect Bandwidth.}
    \label{scale_hw}
\vspace{-15pt} 
\end{figure}

Figure \ref{growth_of_dl} illustrates the exponential growth of computational power for DL. And as summarized in Figure \ref{scale_hw}\cite{gholami_ai_2024}, while AI's demand for computational power is growing at 10x per year, Moore's Law lags behind with hardware FLOPs increase at only 3.0x  every two years, DRAM bandwidth at 1.6x, and interconnect bandwidth at 1.4x. This disparity necessitates more machines, raising DL training costs, particularly for LLM training, where the computational power required surpasses that of traditional HPC applications.

\subsection{Challenges and Solutions in Models Training}

In Deep Learning training, a single task demands hundreds of GPUs and consumes substantial storage and network resources. This massive scale introduces system-level challenges:
\subsubsection{Efficiency}
Firstly, achieving efficient training at this magnitude is crucial. Model FLOPs Utilization (MFU), which assesses the ratio of observed throughput to theoretical maximum throughput (assuming 100\% peak FLOPS), serves as the standard metric for evaluating training efficiency. Training LLMs involves dividing models among GPUs that communicate extensively for progress. Besides communication, factors like operation optimization, data pre-processing, and GPU memory consumption significantly influence MFU. Multiple parallel strategies are employed to enhance efficiency:

\begin{itemize}
\item \textbf{Data Parallelism (DP):} Models and optimizer states are replicated across multiple devices with data evenly distributed to all. For LLMs training, the Zero Redundancy Optimizer (ZeRO)\cite{rajbhandari2020zero} further enhances this method by sharding these states on each data parallel process and using allgather and reduce-scatter for parameter fetching and gradient calculation.

\item \textbf{Pipeline Parallelism (PP):} Each device holds a portion of the model layers with each training batch divided into micro-batches for pipeline execution. Efficient scheduling strategies like GPipe\cite{DBLP:journals/corr/abs-1811-06965}, PipeDream 1F1B\cite{10.1145/3341301.3359646}, and Zero Bubble Pipeline Parallelism (ZBPP)\cite{qi2023zero}, are required to minimize ``pipeline bubbles".
\item \textbf{Tensor Parallelism (TP):} This involves placing a model layer on multiple GPUs that perform computations in parallel\cite{shoeybi_megatron-lm_2020}\cite{10.1145/3458817.3476209}. It includes row-wise and column-wise parallelism, necessitating allgather and all2all for input splitting and output merging.

\item \textbf{Expert Parallelism (EP)}: MoE Models' different expert models are distributed on different GPUs during MoE training\cite{10.1145/3577193.3593704,ae449111733a42c5980594f9133812c8,hwang2023tutel}. The gate model selects tokens for allocation during input, with corresponding tokens sent to experts model via all2all communication.

\item \textbf{Fully Sharded Data Parallel (FSDP)}\label{fsdp_section}  is an implementation based on the ZeRO Stage 3 algorithm\cite{rajbhandari2020zero}. FSDP partitions the model's parameters, optimizer states, and gradients, distributing them across different GPUs, with each GPU retaining only \(1/n\) of the total. During forward propagation, FSDP performs an allgather operation to assemble the complete parameters, which are then released after the forward pass is completed. Similarly, during backward propagation, FSDP conducts an allgather operation to obtain the complete parameters, followed by backward computation to calculate gradients. It then performs a reduce-scatter operation to synchronize gradients across all GPUs, resulting in each GPU holding \(1/n\) of the reduced gradients. Finally, FSDP updates the \(1/n\) parameters using each GPU's \(1/n\) gradients and optimizer states. FSDP reduces GPU memory usage by maintaining only \(1/n\) of the parameters, gradients, and optimizer states on each GPU, enabling training of larger-scale models.
\end{itemize}
There are additional strategies and algorithms to accelerate training or reduce memory usage, such as \textbf{Activation Recomputation}\cite{KorthikantiCLMA23,10.1145/3620666.3651359}, as well as enhanced communication and computation overlap during parallelism\cite{10.1007/978-981-97-0834-5_20}, among others.

\subsubsection{Stability}
The second challenge is achieving high-stability training at scale, i.e., maintaining efficient training throughout the process. Stability is vital from a production standpoint as training a big model with a trillion tokens may span several weeks. In DL training, stragglers and hardware failures are common occurrences rather than outliers. Stragglers can decelerate tasks involving hundreds of GPUs, emphasizing the importance of stability and task recovery time.
\subsection{HPC and AI Clusters of This Era}

\subsubsection{HPC Inadequacies for AI Training}
Traditional supercomputers such as TianHe-2A\cite{dongarra_report_2017}, Stampede 2\cite{Stampede_2}, and Sunway TaihuLight\cite{fu_sunway_2016} primarily focus on double precision calculations and do not support the FP16 precision, rendering them unsuitable for DL training. Fugaku\cite{shimizu_supercomputer_2020}, despite its high performance, does not support tensor GEMM acceleration, a key component for DL workloads.  Although these supercomputers may not be well-suited for DL training, their robust high-performance networks and extensive experience in large-scale cluster construction offer valuable insights and lessons for subsequent researchers. 

\subsubsection{GPU based HPC}
Supercomputers like Frontier\cite{schneider_exascale_2022}, Aurora\cite{Aurora}, Summit\cite{stunkel_high-speed_2020} and Perlmutter\cite{10.1145/3538712.3538734} utilize high-performance GPUs to tackle large-scale computations.  It's worth mentioning that Perlmutter utilizes an all-flash storage system, achieving a peak bandwidth of 5TB/s. Indeed, conducting DL training on these GPU-based HPCs yields significant performance.
\subsubsection{GPU Clusters of Large Companies}
Meta, formerly Facebook, has developed its AI-HPC using a software-hardware co-design approach, with one system utilizing IB and another employing RoCE\cite{mudigere_software-hardware_2023-1,10.1145/3651890.3672233}.  ByteDance initially implemented a DL cluster with a mix of CPU and PCIe GPU\cite{jiang_unied_nodate}. However, with the advent of the LLMs era, they adopted an architecture similar to DGX, building a cluster with over 10,000 GPUs\cite{jiang_megascale_2024}. Alibaba has developed its own HPN network\cite{10.1145/3651890.3672265} for LLMs training using NVIDIA H800 GPUs.  NVIDIA also has its own AI-HPC Eos\cite{NVIDIA_Eos}, which will feature 576 DGX H100 systems totaling 4,608 H100 GPUs. While this will provide a considerable boost in computational power for AI tasks, the high cost of DGX systems raises questions about economic viability.

\subsubsection{AI DSA Clusters}
Custom-designed AI DSA (Domain Specific Architecture) accelerators like Google's TPU\cite{jouppi_tpu_2023} , utilize highly advanced optical switch reconfigurable networks. Alternatives to traditional GPU setups, like Intel Habana Gaudi\cite{zhang_benchmarking_2023}, are also available. Teslahas introduced the Dojo\cite{9895534,10078146} supercomputer , which uses System on Wafer technology to build an entire silicon wafer as a single chip. Huawei has designed the Ascend AI DSA chip\cite{8875654, 9407221}, which remains competitive with NVIDIA, as noted by NVIDIA CEO Jensen Huang. These accelerators are tailored for efficient execution of AI workloads, offering specialized features to optimize model training and inference. However, their software ecosystems, while progressing, still require further development to match the maturity of NVIDIA's offerings.

\subsubsection{Cloud Service Providers}
Cloud service providers, such as Azure, offer flexible and scalable resources for AI training. Despite their convenience and easy accessibility, the costs can accumulate significantly over time. For long-term projects spanning around two years,  these costs could amount to purchasing an entire dedicated cluster. Therefore, this option may not be the most economical choice for extensive AI computations.

\begin{figure}[tb]
    \centerline{}\includegraphics[width=1\linewidth]{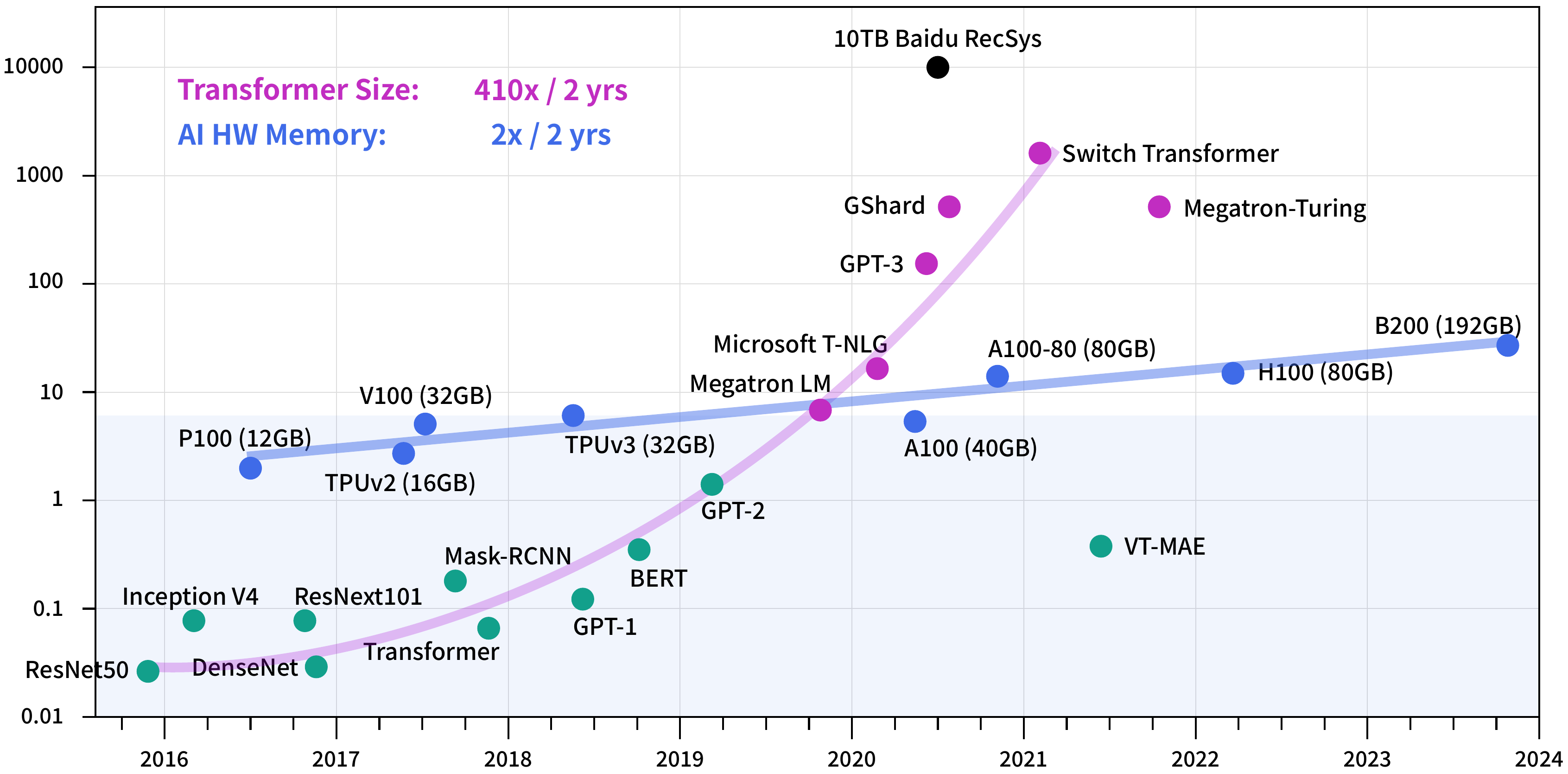}
    \caption{Size of Model Parameter and Accelerator Memory}
    \label{ai_and_memory_wall}
\vspace{-10pt} 
\end{figure}

\subsection{Challenges in AI Infrastructure}
As models continue to grow larger, DL training requires thousands of GPUs. Additionally, researchers often need to train multiple models simultaneously. Therefore, a cluster with at least tens of thousands of GPUs can meet the needs of AI practitioners. In addition to increasing the scale of the cluster nodes and adding more GPUs, there's also a need to find ways to save on the overall system construction costs. These costs include but are not limited to power support, cooling, networking, storage, fault handling, disaster recovery, etc. Building a cost-effective AI-HPC system is a significant challenge. The question of how to construct a high-performance, efficient, economical, and environmentally friendly HPC to meet AI training requirements has become a hot topic. Some works, such as \cite{9474063}  analyzes the construction of HPC, discussed the interconnection of heterogeneous clusters, cooling systems. AI applications such as Peng Cheng Cloud Brain II\cite{su_computing_2021} discussed their strategies for building and improving AI computing power and cluster communication efficiency, which helped them achieve first place in IO500 and AIPerf rankings.

Drawing from our extensive experience in Deep Learning spanning over the past decade, we have conducted considerable exploration in terms of cost-effectiveness. This work primarily discusses our practices for achieving cost-effectiveness and high-performance across different models and stages.

\section{Fire-Flyer 2: Our Approach For Deep Learning and Early LLM Training}
As mentioned in the Background Section, LLMs generally require significant memory resources. In contrast, many other models necessitate considerably less memory, as illustrated in Figure \ref{ai_and_memory_wall}. Popular models like ResNet\cite{he2016deep}, Mask-RCNN\cite{hassan_review_2022}, BERT\cite{Devlin2019BERTPO}, MAE\cite{MaskedAutoencoders2021},  among others, all have a parameter volume less than 1B, signifying relatively low memory requirements. Therefore, when designing a cluster primarily for deep learning model training, and with insights gleaned from our Fire-Flyer 1 experiments, we deemed it prudent to incorporate PCIe A100 GPUs, which proved to be sufficient during its construction in 2021.

\subsection {Fire-Flyer 2: PCIe A100 GPU Architecture}\label{fireflyer2}

\begin{figure}[t]
    \centering
    \includegraphics[width=1\linewidth]{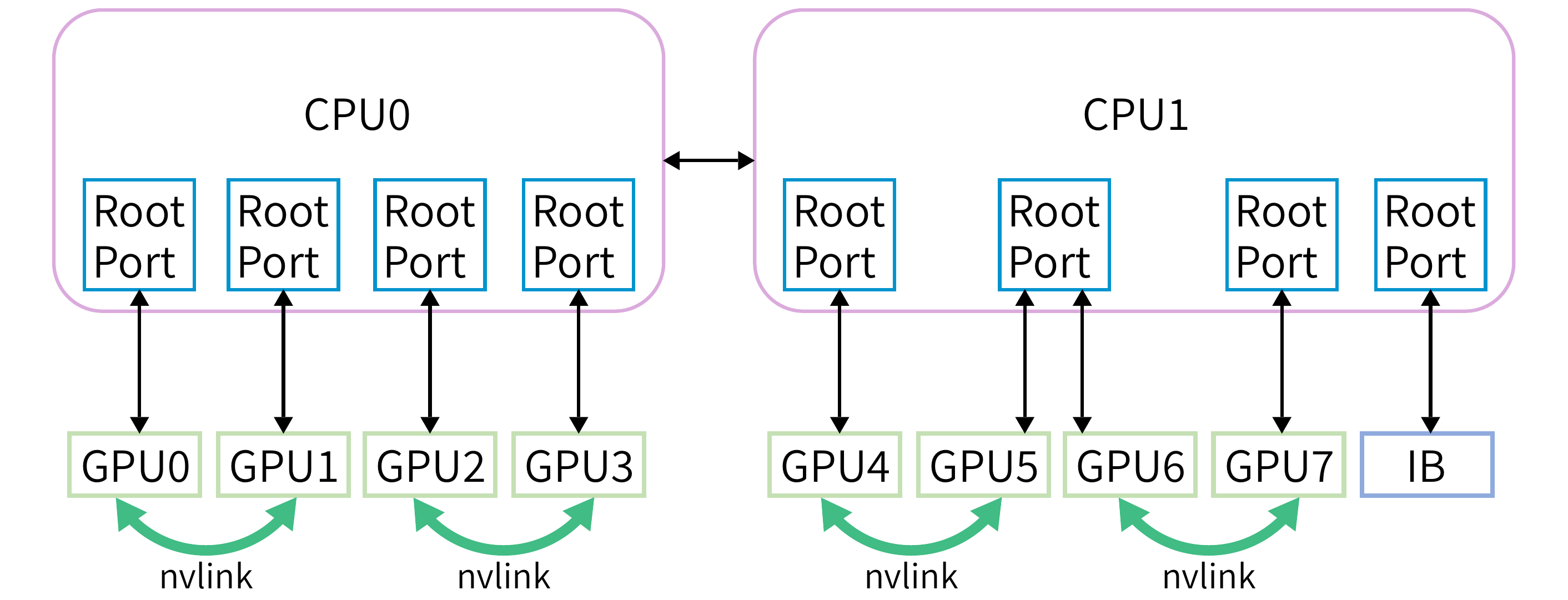}
    \caption{In-node Architecture: 8 PCIe GPUs and 1 InfiniBand (IB) NIC are directly connected to the CPU. Note that GPU5/6 share the same PCIe root port, while IB occupies one independently.}
    \label{node_arch}
\end{figure}

In our training workloads, the bandwidth requirements for both storage IOs and computation communication across 8 NVIDIA PCIe A100 GPUs can be met by a single 200Gbps NVIDIA Mellanox ConnectX-6 (CX6) InfiniBand (IB) NIC. We employed the following computation node architecture , as shown in Figure \ref{node_arch}:\

\begin{table}[tb]
\centering
\caption{Our Arch and DGX-A100 Server Hardware Details}
\setlength{\tabcolsep}{10pt} 
\renewcommand{\arraystretch}{1.5} 

\newcolumntype{Y}{>{\raggedright\arraybackslash}X}
\newcolumntype{Z}{>{\raggedright\arraybackslash}p{.5\linewidth}}

\begin{tabularx}{0.5\textwidth}{|l|Y|Y|}
\hline
        & Our PCIe Arch             & DGX-A100                     \\
\hline
CPU    & 2 * AMD 32 Cores EPYC Rome/Milan CPU & 2 * AMD 64 Cores EPYC 7742 CPU  \\
\hline
Memory & 512GB 16-Channels DDR4-3200Mhz & 2048GB 16-Channels DDR4-3200Mhz    \\
\hline
GPU    & 8 * PCIe-A100-40GB            & 8 * SXM-A100-40GB                  \\
\hline
NICs   & 1 * Mellanox InfiniBand cx6 200Gbps NIC & 9 * Mellanox InfiniBand cx6 200Gbps NIC \\
\hline
NVLINK & 600 GB/s between each pair of GPUs & 600 GB/s interconnect among all 8 GPUs \\
\hline
\end{tabularx}
\label{arch_details}
\end{table}

\begin{itemize}
\item 8 NVIDIA A100 PCIe GPUs and 1 Mellanox CX6 200Gbps IB NIC: directly connect to the CPU, without using a PCIe switch
\item IB NIC occupies a separate PCIe root complex, thus avoiding performance interference with the GPU.
\item  Reserved the possibility of NVLink Bridge addition in design: As expected, when the LLM era arrived, we indeed added an NVLink Bridge between PCIe cards.
\end{itemize}

Table \ref{arch_details} shows our arch details and compared with NVIDIA standard DGX-A100 server.

\subsection {Network Topology: Two-Layer Fat-Tree with Storage and Computation Integrated}\label{network_topo_section} 

We selected the \textbf{Fat-Tree}\cite{fat_tree_6312192} topology as our primary network architecture due to its exceptionally high bisection bandwidth, making it the preferred choice for AI-HPC and  high-throughput storage environments. Although the Dragonfly topology\cite{4556717,10.1145/3524059.3532380} also offers comparable cost-effectiveness and performance,  its lack of sufficient bisection bandwidth makes it  unsuitable  for our \textbf{integrated storage and computation network design}.  At the time of implementation, various RoCE (RDMA over Converged Ethernet) \cite{5289144} technologies were not as mature as they are today, so we opted for InfiniBand (IB) as our network solution. Mellanox QM8700 InfiniBand Switch,  offering 40 ports at 200 Gbps, was utilized. Our cluster, consisting of 10,000 A100 GPUs, includes approximately 1,250 GPU compute nodes and nearly 200 storage servers,  although a Two-Layer Fat-Tree can accommodate up to 800 nodes (configured  with 20 spine switches and 40 leaf switches).

\begin{figure}[bp]
    \centering
    \includegraphics[width=1\linewidth]{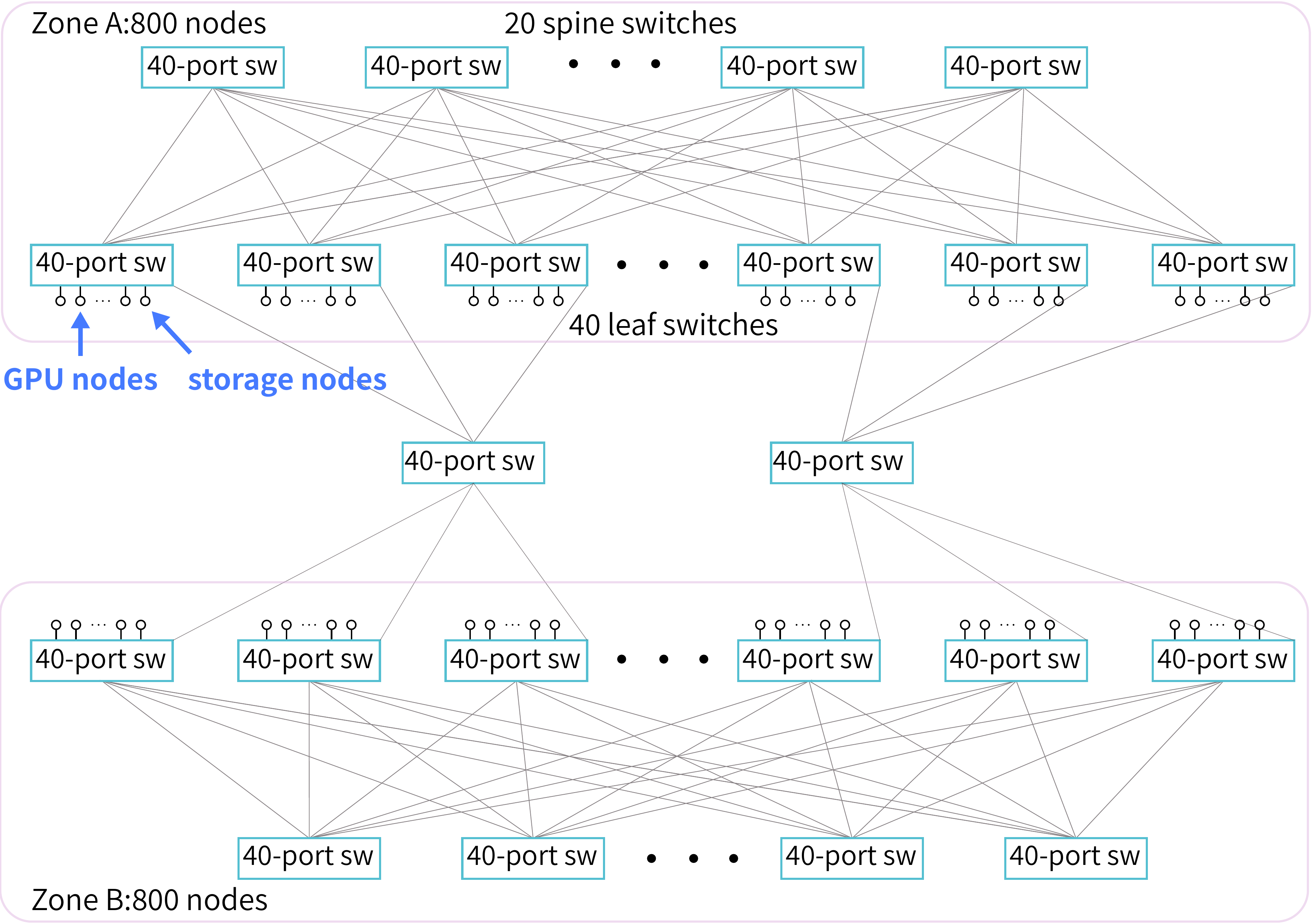}
    \caption{Network Topology: Two complete Two-Layer Fat-Tree connected together.}
    \label{net_topo}
\end{figure}

To reduce costs, , we opted for a two-zone network configuration instead of a three-layer Fat-Tree solution, as shown in Figure \ref{net_topo}. Each zone consists of an 800-port Fat-Tree connected to approximately 600 GPU compute nodes. Each storage server equipped with two IB NICs, respectively connected to different zones, hence all GPU compute nodes could share a set of storage services. Additionally, the two zones are interconnected with a limited number of links. Our HAI Platform scheduling strategy ensured that cross-zone computing tasks were limited to one at most. Whether using NCCL\cite{NCCL_LINK} or our in-house developed communication library HFReduce, it can be run across zones by using a double binary tree algorithm\cite{double_binary_tree}.  Our scheduler ensures that in this topology, only one pair of nodes communicates across zones. Consequently, even tasks requiring all nodes can efficiently run on the entire Fire-Flyer 2 AI-HPC.

\subsection{Cost Performance of Our Architecture}
Compared to the NVIDIA DGX-A100 \cite{nvidia_dgx} architecture, our approach using PCIe A100 achieves approximately 83\% of the performance in TF32 and FP16 General Matrix Multiply (GEMM) benchmarks. However, it offers substantial reductions in both costs and energy usage, achieving 60\% of the GPU cost and energy consumption, as detailed in Table \ref{pcie_cost}.

Contrasting with the DGX-A100 cluster, which necessitates a Three-Layer Fat-Tree encompassing 10,000 access points and involving 320 core switches, 500 spine switches and 500 leaf switches, amounting to 1,320 switches in total (as shown in Table \ref{cost_compare}), our architecture only requires 122 switches. This arrangement is significantly more cost-efficient. Even when compared to a similarly sized three-layer Fat-Tree network with 1,600 access points, which includes 40 core switches and 160 spine and leaf switches (totaling 200 switches), our design facilitates a saving of 40\% in networking costs.

Furthermore, by utilizing an 800-Ports Frame Switch, we have further reduced the cost of optical modules and cables. While there is a performance gap due to the inherent differences between PCIe card specifications and SXM, we generally achieved 80\% of the DGX-A100 performance at merely 60\% of the cost. Additionally, we managed to trim energy consumption by 40\%, thereby reducing CO\textsubscript{2} emissions. In terms of cost-performance, we regard this approach as both effective and successful.

\begin{table}[tb]
\centering
\caption{A100 PCIe Compared to DGX-A100.}
\setlength{\tabcolsep}{10pt} 
\renewcommand{\arraystretch}{1.25} 

\newcolumntype{Y}{>{\centering\arraybackslash}X}
\newcolumntype{Z}{>{\centering\arraybackslash}p{.5\linewidth}}

\begin{tabularx}{\linewidth}{|Z|Y|Y|Y|}
\hline
\multicolumn{2}{|c|}{} & Our Arch & DGX Arch \\
\hline
\multicolumn{2}{|c|}{TF32 GEMM (TFLOPS/GPU)} & 107 &  131 \\
\hline
\multicolumn{2}{|c|}{FP16 GEMM (TFLOPS/GPU)} & 220  & 263 \\
\hline
\multicolumn{2}{|c|}{Relative Performance} & 83\% & 100\% \\
\hline
\multicolumn{2}{|c|}{Node Relative Price} & 60\% & 100\%  \\
\hline
\multicolumn{2}{|c|}{Cost–Performance Ratio} & 1.38 & 1  \\
\hline
\multicolumn{2}{|c|}{Power Consumption (Watts)} & 2500 & 4200 \\
\hline
\end{tabularx}
\label{pcie_cost}
\end{table}

\begin{table}[tp]
\centering
\caption{Relative Cost Comparison.}
\setlength{\tabcolsep}{10pt} 
\renewcommand{\arraystretch}{1.25}

\newcolumntype{Y}{>{\centering\arraybackslash}X}
\newcolumntype{Z}{>{\centering\arraybackslash}p{.28\linewidth}}

\begin{tabularx}{\linewidth}{|l|X|X|Z|X|}
\hline
\multicolumn{2}{|c|}{} & Our Arch & PCIe Arch with Three Layer Fat-Tree & DGX Arch \\
\hline
\multicolumn{2}{|c|}{Number of Switches} & 122 & 200 & 1320 \\
\hline
\multicolumn{2}{|c|}{Network Price} & 350 & 600 & 4000 \\
\hline
\multicolumn{2}{|c|}{Server Price} & 11250 & 11250 & 19000 \\
\hline
\multicolumn{2}{|c|}{Total Price} & 11600 & 11850 & 23000 \\
\hline
\end{tabularx}
\label{cost_compare}
\vspace{-5pt} 
\end{table}

\section{HFReduce: Hardware Software Co-Design in Network}\label{HFReduce_section}
\begin{figure}[tbp]
\centerline{\includegraphics[width=0.5\textwidth]{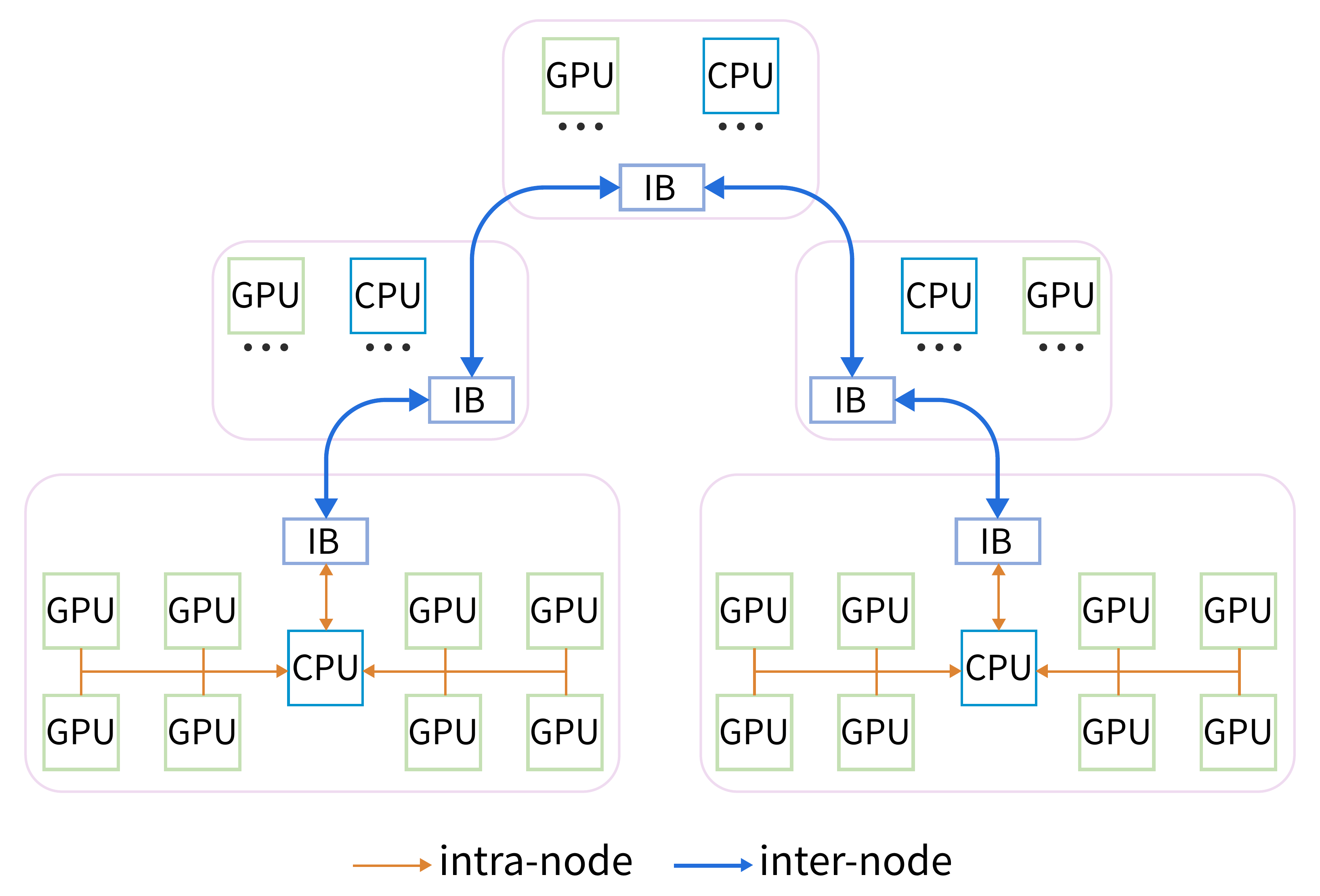}}
\caption{HFReduce Schematic: 1) do intra-node reduce, 2) do inter-node allreduce by CPU, 3) finally transfer reduced data to GPU.}
\label{fg_hfreduce_concept}
\vspace{-15pt} 
\end{figure}

In large-scale deep learning training, allreduce is essential for aggregating gradients across GPUs. To optimize communication among PCIe GPUs in our architecture, we developed HFReduce, a library specifically designed for efficient allreduce operations. The core strategy of HFReduce, illustrated in Figure~\ref{fg_hfreduce_concept}, involves performing intra-node reduction first, followed by inter-node allreduce of the reduced data from the 8 GPUs within each node. This inter-node allreduce leverages a Double Binary Tree Algorithm\cite{double_binary_tree}, akin to NCCL, and is pipelined by dividing data into chunks for transfer via Remote Direct Memory Access (RDMA), ensuring high performance. HFReduce is versatile and can be applied to any scenario requiring allreduce, as well as general reduce and broadcast operations.

\subsection{HFReduce Algorithm Steps}

\textbf{Intra-node} reduction, as shown in the Algorithm \ref{algo_hfreduce_intra_node}:
\begin{enumerate}
\item  When the gradients data on the GPUs require allreduce, HFReduce asynchronously transfers these data to the CPU memory. This Device-To-Host (D2H) Transfer can utilize GDRCopy\cite{7116873} for small data and MemCpyAsync for larger data.
\item Upon the arrival of the gradients in memory, perform reduction add operation using CPU vector instructions.
\end{enumerate}

\textbf{Inter-node} reduction, as shown in the Algorithm \ref{algo_hfreduce_inter_node}:
\begin{enumerate}
    \item  Use the Double Binary Tree Algorithm\cite{double_binary_tree} for inter-node allreduce, facilitating transfers between nodes using RDMA verbs implementation.
    \item  Finally, the CPU returns reduced gradients to the GPU via PCIe (Host-To-Device Phase).
\end{enumerate}
The final Host-To-Device (H2D) Transfer can be optimized by utilizing GDRCopy to write data to four GPUs within the same NUMA node, effectively reducing reads from host memory by threefold compared to MemCpyAsync. This efficiency is achieved because GDRCopy can read data from host memory and temporarily cache it in the CPU caches, allowing data to be written to the four GPUs without additional reads from host memory.

\begin{algorithm}[t]
\caption{Intra Node Reduce.}
\label{algo_hfreduce_intra_node}
    \KwData{Dg: data need to allreduce}
\KwResult{Dc: data reduced in this node}
Split Dg by Chunk\_Size\\
\For {Dg-i in Splited-Dg}{
    \tcp{Transfer GPU Memory Data to CPU memory}
    Dc\_i = MemCopyAsync Dg\_i to CPU Mmeory\\
}
\For {i in Splited\_Count}{
\While{every GPU's $Dg\_i\rightarrow Dc\_i$\   finished}{
\tcp{Wait for chunk-i transfer finished in this node}
    \For {j in GPU\_Count}{
       
        \tcp{do intra-node reduce}
       Dc\_i += GPU-j's Dc\_i
    }
        \tcp{do inter-node reduce}
    \textbf{Do\_Internode\_reduce(Dc\_i)}
}
}
\end{algorithm}

\begin{algorithm}[bt]
\caption{Inter Node Reduce.}
\label{algo_hfreduce_inter_node}
\KwData{DL:local node reduced data by Algorithm 1}
\KwData{DR: received other node reduced data}
\KwResult{Dg: reduced data transfer to GPU}
\tcp{Pass 1: reduce data,individual thread}
\For {i in Chunk\_Size}{
receive data $DR\_i$ from prev node\\
\If{$DL\_i$ ready}{
       DL\_i +=DR\_i \tcp{reduce data}
}
\uIf{Thread is root of Tree}{
    
    send DL\_i to prev node \tcp {Dg\_i finished, go parse 2}
}\Else{
    send DL\_i to next node
}
}
\tcp{}
\tcp{Pass 2: gather reduced data,individual thread}
receive data $DR\_i$ from next node\\
\For {j in GPU\_Count}{
    Dg\_i = MemCopyAsync DR\_i to GPU-j \tcp{Dg\_i is allreduced}
}
send DL\_i to prev node
\end{algorithm}

\subsection {Advantages of HFReduce over NCCL}
 \subsubsection{\textbf{Reduced PCIe Bandwidth Consumption}} Let $n$ be the total number of GPUs involved in the communication. In NCCL's ring topology, each unit of data needs to go through $2n-1$ transmissions, each consuming one unit of inbound bandwidth of one GPU and one unit of outbound bandwidth of another GPU. This means for a single unit of data, it consumes $\frac{2n-1}{n}$  unit of PCIe bidirectional bandwidth. In contrast, for each unit of data, HFReduce only requires one D2H and one H2D data transfer, only one unit of PCIe bidirectional bandwidth is consumed. In our machine architecture, the performance of NCCL is mainly limited by PCIe bandwidth. Therefore, HFReduce can achieve better performance than NCCL.
 
\subsubsection{\textbf{No GPU Kernel Overhead}} HFReduce utilizes the GPU's Copy Engine (CE) for PCIe asynchronous transfers. In contrast, NCCL's allreduce operation requires GPU kernel execution, which can affect other computational kernels on the GPU. HFReduce achieves complete asynchrony with no overhead. 

\begin{figure*}[ht]
    \centering
    \begin{subfigure}{0.49\textwidth}
        \includegraphics[width=\linewidth]{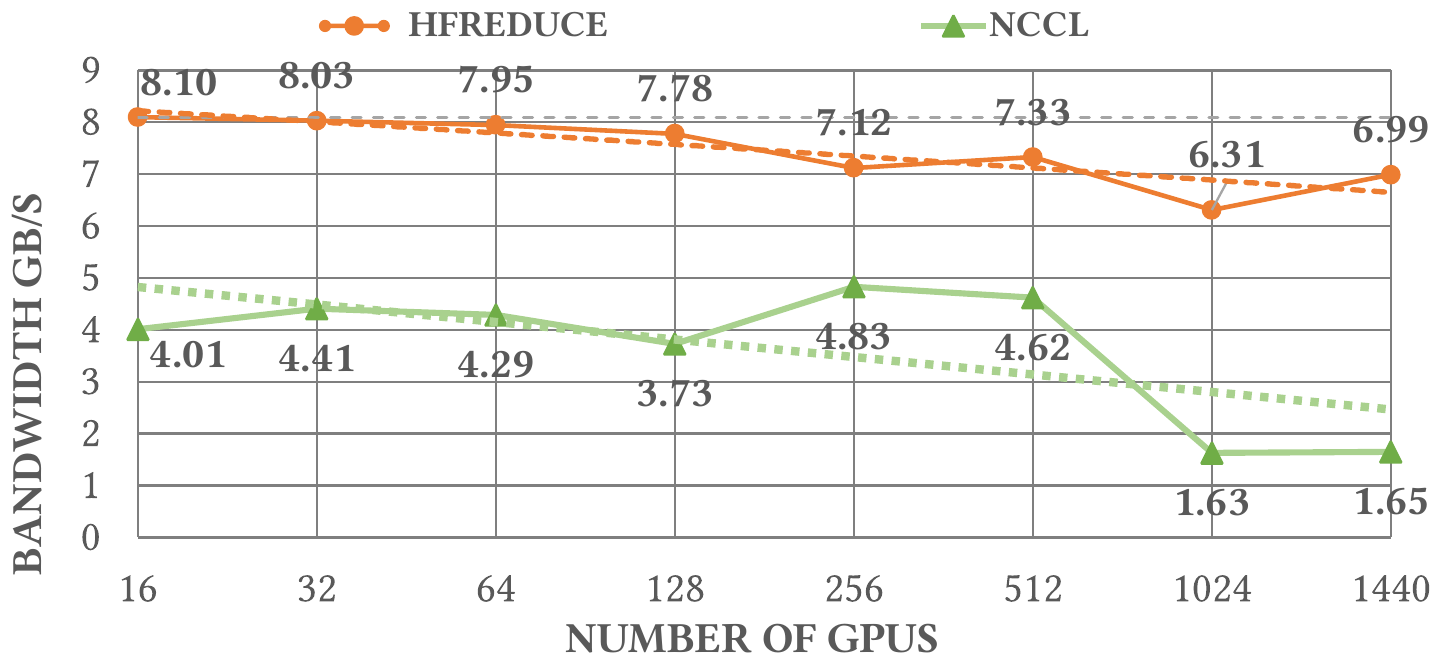}
        \caption{HFRudce and NCCL Allreduce Speed.}
        \label{hfreduce_nccl_perf}
    \end{subfigure}\hfill
    \begin{subfigure}{0.51\textwidth}
        \includegraphics[width=\linewidth]{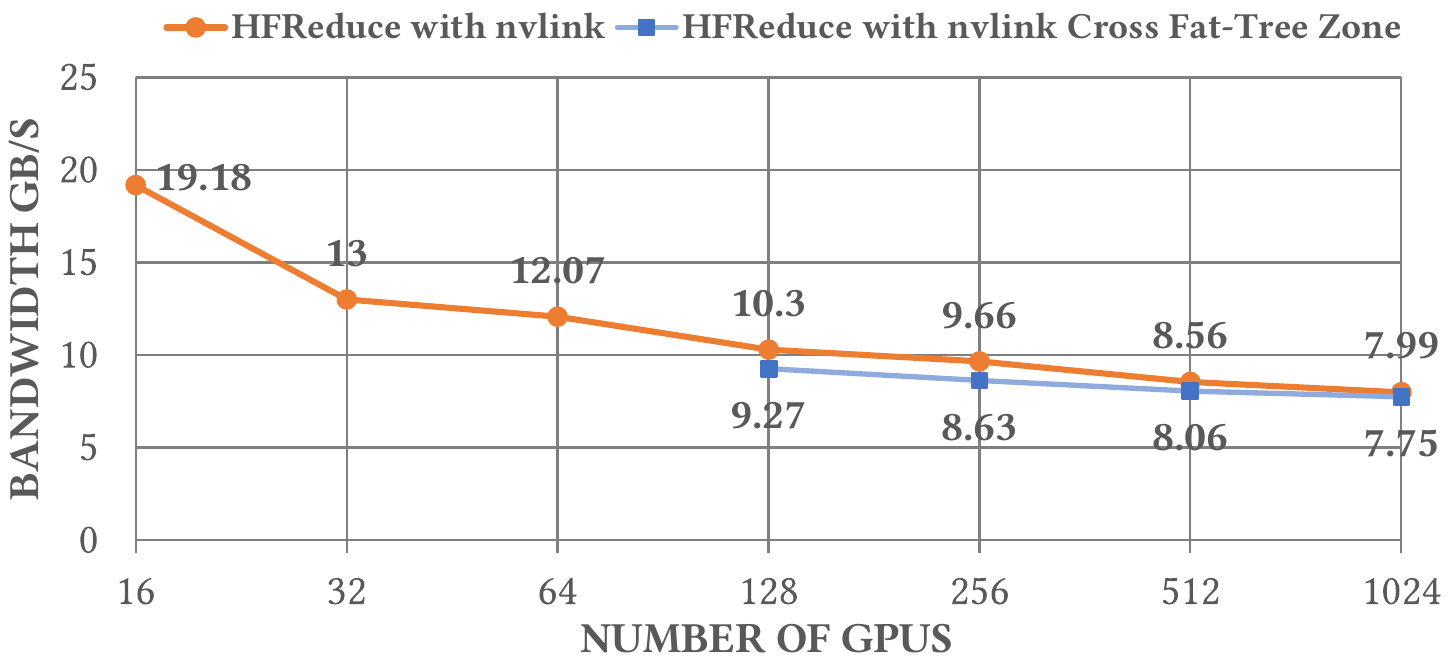}
        \caption{HFReduce with NVLink (Cross Fat-Tree Zone).}
        \label{cross_plane_perf}
    \end{subfigure}\hfill
        \caption{Strong Scalability: (a) Network Bandwidth when HFReduce and NCCL do allreduce test with 186MiB data, scale from 16 to 1440 GPUs. (b) HFReduce with NVLink, and Cross Fat-Tree Zone.  Note that tasks utilizing fewer than 128 GPUs do not require cross-zone nodes and are restricted by platform defaults. }
\vspace{-15pt} 
\end{figure*}

As demonstrated in Figure \ref{hfreduce_nccl_perf}, HFReduce can reach a inter-node bandwidths of 6.3-8.1GB/s when performing allreduce with a data size of 186 MiB on the Fire-Flyer 2 AI-HPC, while NCCL's inter-node bandwidth is only 1.6-4.8GB/s.

\subsection{Performance Improvements: HFReduce with NVLink}

By installing the NVLink Bridge for PCIe A100 GPUs, efficient communication is enabled between paired GPUs via the 600 GB/s NVLink. To alleviate the memory bound issue of the original HFReduce, we implemented another allreduce pattern, termed \textbf{ HFReduce with NVLink}. The core concept involves initially performing a reduction operation among GPUs interconnected by NVLink before passing the gradient to the CPU. Subsequently, when the CPU returns the result, it splits the result data and returns them to the paired GPUs connected by NVLink respectively, then performs allgather via NVLink. As illustrated in Figure \ref{cross_plane_perf}, HFReduce with NVLink achieves inter-node bandwidths exceeding 10 GB/s.

\subsection{Deep Analysis of HFReduce}
\subsubsection{Key Technical Strategies in Implementation}
\begin{itemize}
    \item Using GDRCopy accelerate small data transfer in D2H, and educing reads from host memory by three times compared to MemCpyAsyn.
    \item Intra-Node Reduction: CPU utilizes SIMD instructions and supports FP32 / FP16 / BF16 / FP8 datatypes.
    \item NUMA Awareness: D2H destination memory is interleaved across two NUMA nodes for maximum bandwidth. Memory for CPU-added results and network-received data is bound to the IB-Nic's NUMA node to minimize latency.
    \item Inter-Node Reduce:Implements a Double Binary Tree allreduce algorithm\cite{double_binary_tree} via ibverbs RDMA Write, avoiding additional overhead.
\end{itemize}
\subsubsection{HFReduce Overcomes Limitations of EPYC Rome CPU} We consulted AMD and NVIDIA engineers to identify the root cause of NCCL's suboptimal performance on PCIe architecture, particularly with EPYC Rome CPU servers. It was determined that the Rome CPUs do not support the chained write feature, which can significantly accelerate PCIe peer-to-peer (P2P) transfers between GPUs and IB NICs. Our tests indicate that the maximum bandwidth between the GPU and IB NIC on Rome CPUs is approximately 9 GiB/s, making the observed 4GB/s all-reduce bandwidth for NCCL understandable. HFReduce circumvents this limitation by utilizing the CPU for reduction and transferring data through IB and host memory.

\subsubsection{\textbf{Bottlenecks of HFReduce}}

When considering the total memory operations on a single node during HFReduce, several factors contribute to its performance limitations:

\begin{enumerate}
    \item D2H Phase requires 8 write operations.
    \item  Intra-node Reduce Add Phase involves 8 read operations and 1 write operation.
    \item Inter-node Allreduce Phase: IB send demands 2 read operations, while IB receive requires 2 write operations, along with 1 read operation for reduce add.
    \item H2D Phase Utilizing GDRCopy can reduce this to only 2 read operations, whereas MemCopy necessitates 8 read operations.
\end{enumerate}

\begin{figure}[bt]
\vspace{-20pt} 
    \centering
    \begin{subfigure}{0.2575\textwidth}
        \includegraphics[width=\linewidth]{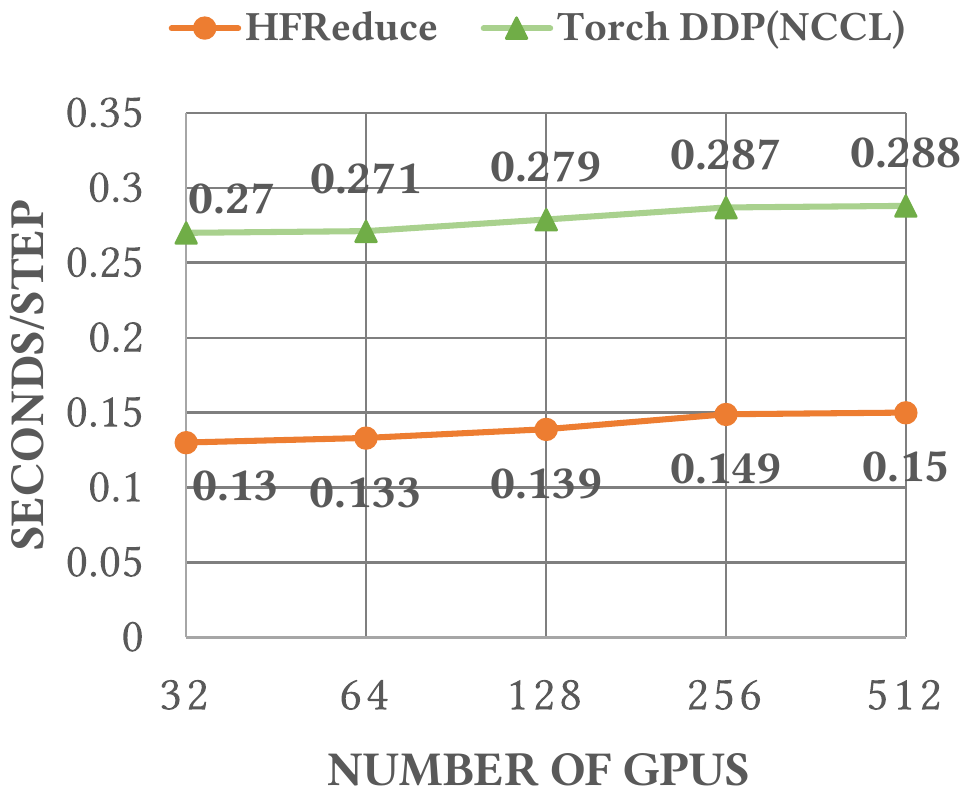}
        \caption{HFReduce v.s. Torch DDP. }
        \label{vgg16_train}
    \end{subfigure}
    \begin{subfigure}{0.2225\textwidth}
        \includegraphics[width=\linewidth]{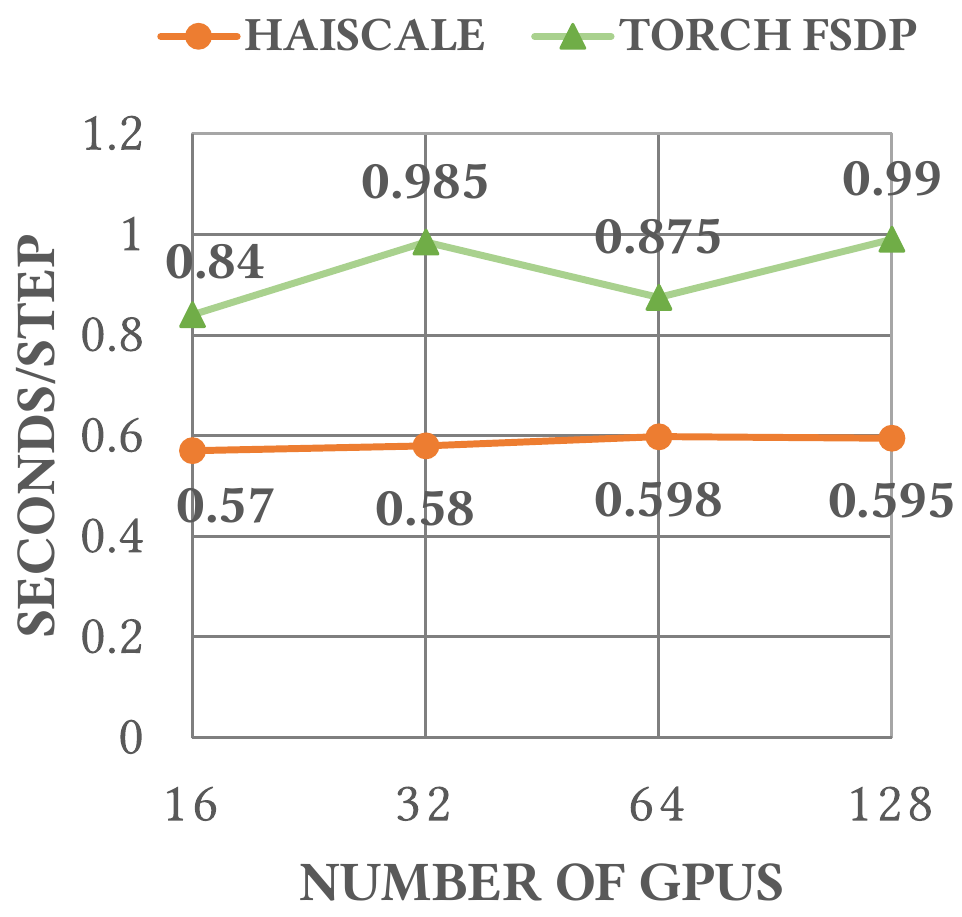}
        \caption{HaiScale v.s. Torch FSDP.}
        \label{haiscale_perf}
    \end{subfigure}
    \caption{Weak Scalability: (a) Training VGG16, HFReduce compared to PyTorch DDP's NCCL backend (b) Training GPT2-Medium HaiScale compared to Torch, both using FSDP.}
\vspace{-10pt} 

\end{figure}

In total, the memory operations amount to 24 times the original data size in the GPU. A host equipped with 16 channels of DDR4-3200MHz memory can achieve a practical memory access speed of 320GB/s. Consequently, the theoretical maximum speed of HFReduce is approximately 13.3GB/s, but when considering the allreduce algorithm bandwidth and network bandwidth, this value realistically approximates 12GB/s. However, our tests only achieved slightly over 8GB/s.

The root cause of this discrepancy is another limitation of the EPYC CPUs. As previously mentioned, our GPU5 and  GPU6 are directly connected to the CPU via the same PCIe Root Complex Port (also known as the PCIe Host Bridge). In AMD EPYC Rome and Milan CPUs, the maximum bandwidth from the Root Complex Port to the CPU's internal bus is about 37.5GB/s. Although a PCIe 4.0 x16 port can achieve over 27GB/s from GPU to CPU, when two GPUs transfer data concurrently, the bandwidth is limited to around 37GB/s. Furthermore, if bidirectional data transfer occurs simultaneously, this bandwidth decreases even further. As a result, HFReduce does not reach its theoretical speed.

Employing NVLink with HFReduce offers a functional method to alleviate these bottlenecks. However, it is worth noting that the next-generation CPUs, such as the EPYC Genoa, still face issues with PCIe Host Bridge bandwidth, which cannot support two full-speed PCIe ports simultaneously. We hope AMD will address this issue in future iterations.

\section{HaiScale: Special Optimization for Deep Leaning Models Training}\label{HaiScale_section}
\subsection{HaiScale DDP Overlap AllReduce in Training}

HaiScale Distributed Data Parallel (DDP)  is a training tool that utilizes HFReduce as its communication backend, in contrast to PyTorch's DDP\cite{pytorch_git} which employs NCCL as its backend.  During the backpropagation phase, HaiScale DDP performs an asynchronous  allreduce operation on the computed gradients, allowing this communication to overlap with the computation involved in backpropagation.

As previously mentioned, HFReduce does not depend on GPU Streaming Multiprocessors (SM) for reduction computation, enabling completely asynchronous allreduce without impacting performance. As shown in Figure \ref{vgg16_train}, training VGG16 model\cite{7486599} with HFReduce takes only half the time compared to using Torch DDP's NCCL backend, achieving nearly 88\% parallel scalability when scale from 32 GPUs to 512.

\subsection{LLMs Training Optimization}
Our HaiScale framework  various parallelism strategies for training large language models (LLMs), similar to Megagron\cite{shoeybi2019megatron} and DeepSpeed\cite{10.1145/3394486.3406703}.  We have made specific engineering optimizations for our PCIe architecture across Data Parallelism (DP), Pipeline Parallelism (PP)\cite{DBLP:journals/corr/abs-1811-06965}\cite{10.1145/3341301.3359646}, Tensor Parallelism (TP)\cite{10.1145/3458817.3476209}, Expert Parallelism (EP) \cite{10.1145/3577193.3593704,ae449111733a42c5980594f9133812c8,hwang2023tutel}.

\subsubsection{NVLink Bridge Enables Tensor Parallel between PCIe GPUs} 
With the advent of LLMs, we integrated the NVLink Bridge into our system. This addition established a bandwidth of 600GB/s between each pair of GPUs, enabling more efficient when performing Tensor Parallelism.

\subsubsection{Pipeline Parallelism Optimization in PCIe Architecture}
\begin{figure}[tb]
\vspace{-20pt} 
    \centering
    \begin{subfigure}{0.2125
    \textwidth}
        \includegraphics[width=\linewidth]{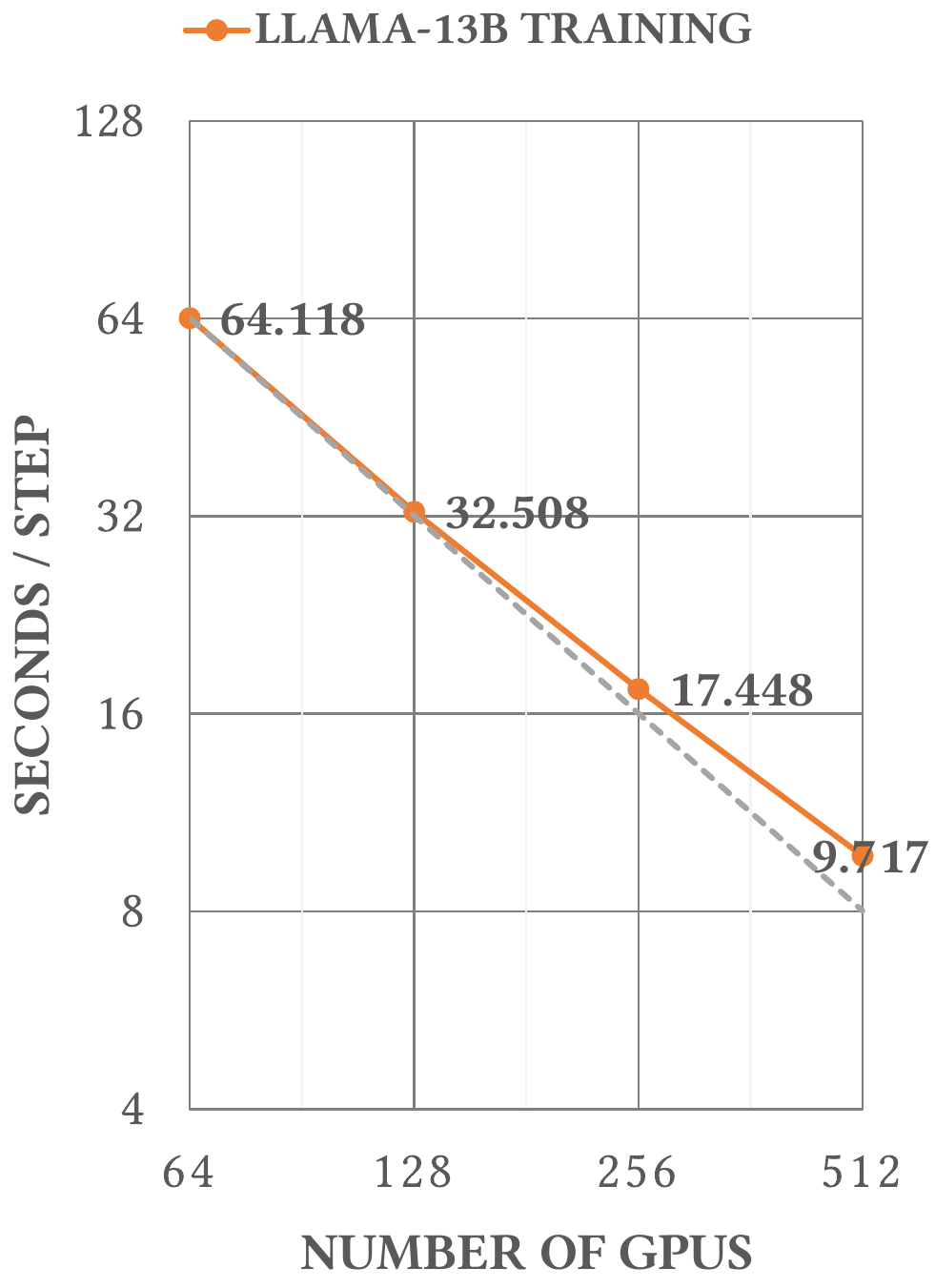}
        \caption{Train LLaMa-13B}
        \label{llama_13b_train}
    \end{subfigure}
    \begin{subfigure}{0.2675\textwidth}
        \includegraphics[width=\linewidth]{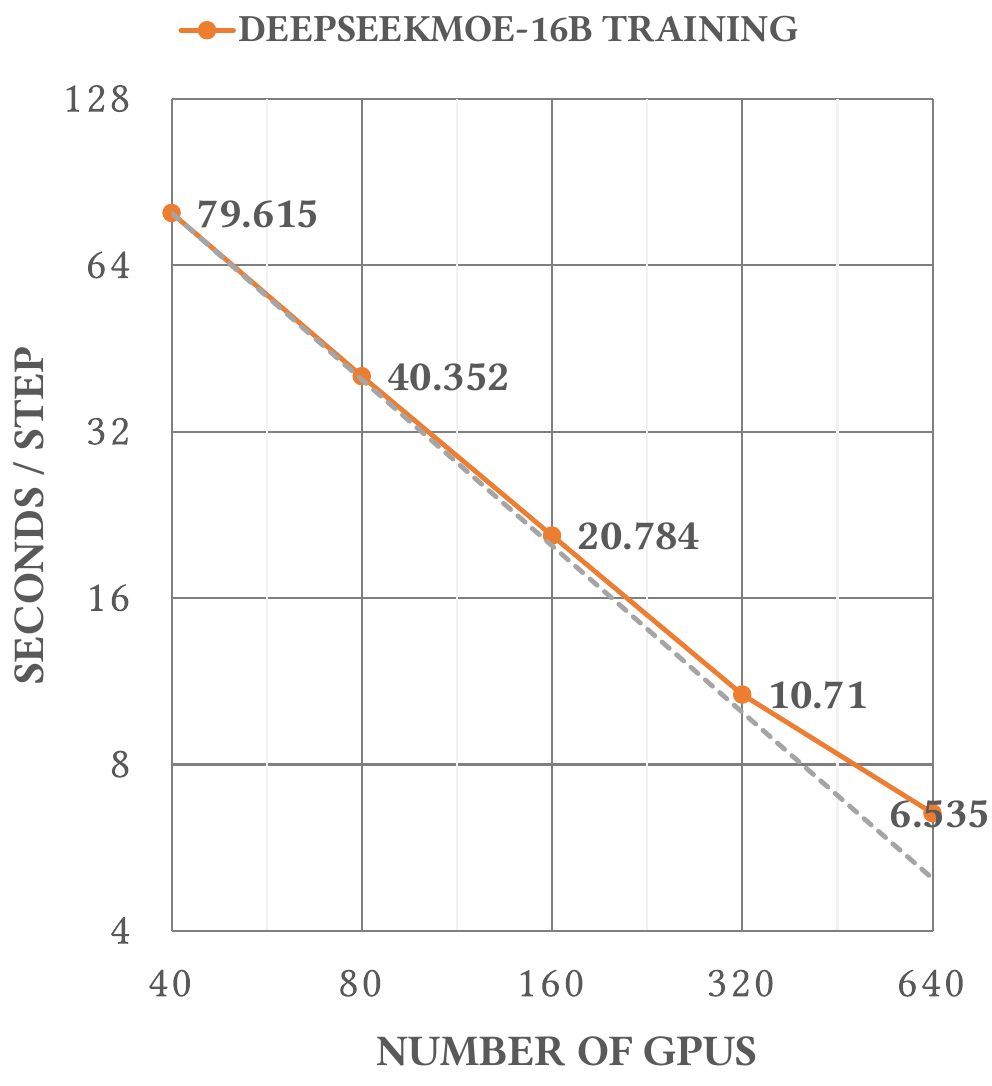}
        \caption{Train DeepSeekMoE-16B}
        \label{moe_16b_train}
    \end{subfigure}
    \caption{Strong Scalability: (a) Train LLaMa-13B with a config of sequence length 2048, batch size 4096, pipeline parallel 4. (b)Train DeepSeekMoE-16B with a config of sequence length 4096, batch size 4608, pipeline parallel 10.}
\vspace{-10pt} 
\end{figure}

In our architecture, there is only one IB NIC for 8 GPUs on a single node, which can lead to network bandwidth contention during Pipeline Parallelism (PP). We solve this by configuring Data Parallelism (DP) rank, making the 8 GPUs on the same node belong to different DP ranks which staggers the timing of PP for each DP rank. As Figure \ref{llama_13b_train} shown, when scaling from 64 GPUs to 512 GPUs, the step time of training LLaMa-13B\cite{Touvron2023LLaMAOA} decreases from 64.118 seconds to 9.717 seconds, achieving a parallel efficiency of \textbf{91\%}.

We also benchmarked the training performance of our DeepSeekMoE-16B model \cite{deepseek-ai_deepseek-v2_2024} on Fire-Flyer 2 AI-HPC. As shown in Figure \ref{moe_16b_train}, scaling from 40 GPUs to 640 GPUs reduced the time per training step from 79.615 seconds to 6.535 seconds, achieving a parallel efficiency of 76.14\%. Notably,  with 320 GPUs, the step time was 10.71 seconds, resulting in a parallel efficiency of \textbf{92.92\%}, demonstrating excellent scalability.

\subsubsection {Fully Sharded Data Parallel (FSDP)} 
Both HaiScale's FSDP and PyTorch's FSDP\cite{zhao2023pytorch}  are implementations based on the ZeRO Stage-3 algorithm\cite{rajbhandari2020zero}. The details of this implementation are already discussed in Section \ref{fsdp_section}.

HaiScale's FSDP offers better engineering implementation, optimizing memory management to reduce fragmentation specific to model adjustments. And we overlap allgather and reduce-scatter communication with forward and backward computation, split the optimization step during backward propagation for enhanced overlap. As shown in Figure \ref{haiscale_perf}, training GPT2-medium\cite{radford2019language}, we achieve 95\% parallel scalability when scaling from 16 to 128 GPUs. ompared to PyTorch's FSDP, HaiScale's FSDP reduces training time by nearly half.

\subsection{Summary}
Our AI-HPC design meets DL requirements, and with the addition of the NVLink Bridge, it meets the training needs of early-stage LLMs, reaching the utilization upper limit of PCIe GPUs. However, due to the inherent gap between PCIe card specifications and SXM, there is a certain performance discrepancy. Considering overall performance, basic setup cost, and energy consumption, we achieved 80\% performance at half the cost. We believe Fire-Flyer 2 AI-HPC is a successful practice in terms of cost-effectiveness.

\section{Advanced Cost-Effective and Co-Design Optimizations}

\subsection{Ensuring Minimal Congestion in Our Computation-Storage Integrated Network}\label{network_tuning_section}

As previously stated, our cost-effective network integrated computation communication and storage traffics together. To achieve maximum bandwidth, it is essential to isolate interference between different types of traffic and control network congestion. In practice, we implemented the following measures:
\subsubsection{Divergence of Different Traffics}
In typical training tasks, there are four different types of traffic: HFReduce communication, NCCL communication, 3FS storage traffic, and other traffic. By using InfiniBand's Service Level (SL) technology\cite{1668378}\cite{iba_qos}, we assign different value of SL when establishing connections between nodes and map SL to IB physical queues Virtual Lanes (VLs)\cite{1668378}\cite{iba_qos}. The use of  Virtual Lanes ensures that flows in distinct lanes do not interfere with each other. Ultimately, we configured their proportions to implement traffic isolation, thereby preventing network congestion caused by Head-of-line (HOL) blocking \cite{4150963} and different traffic collisions.

\subsubsection{Topology Adjustment and Route Optimization}
In high-throughput storage scenarios, there naturally exist many incast communication patterns, leading to certain congestion in the network. Under such circumstances, we observed that enabling adaptive routing would lead to more severe congestion spread in the network. Therefore, we opted for a static routing strategy. Based on the static routing scheme, to evenly disperse storage traffic into leaf $\rightarrow$ spine links, we distribute various nodes (storage, computation, management nodes) evenly disperse storage traffic into leaf $\rightarrow$ spine links.

\subsubsection{NCCL Optimization}
We adjusted the NCCL topology to route through the IB NIC and GPUs within the same NUMA node. This adjustment reduced PCIe congestion caused by CPU chiplet interconnects. Additionally, by using PCIe Relaxed Ordering\cite{10.5555/861280}, we further reduced congestion and increased bandwidth.

\subsubsection{Network Tuning in 3FS}
3FS implements  a request-to-send control mechanism to mitigate the congestion. Details are discussed in the next subsection, \nameref{3fs_key_points}.

\subsection{\textbf{High-Throughput Distributed File System: 3FS}}\label{3fs_section}

\subsubsection{Overview}
3FS is our in-house developed high performance distributed file system, akin to WekaFS\cite{wekafs}, DAOS\cite{10.1007/978-3-030-48842-0_3, 10.1145/3581576.3581577}, and BeeGFS\cite{beegfs}. However, the design and implementation of 3FS specifically focus on fully utilizing the high IOPS and throughput of NVMe SSDs and the RDMA network. 

\begin{table}[ht]
\centering
\caption{Storage Node Hardware Details }
\setlength{\tabcolsep}{10pt} 
\renewcommand{\arraystretch}{1.5} 

\newcolumntype{Y}{>{\raggedright\arraybackslash}X}
\newcolumntype{Z}{>{\raggedright\arraybackslash}p{.5\linewidth}}

\begin{tabularx}{0.5\textwidth}{|l|X|}
\hline
CPU    & 1 * AMD 64 Cores EPYC 7742 CPU  \\
\hline
Memory & 512GB 8-Channels DDR4-3200Mhz   \\
\hline
NICs   & 2 * Mellanox InfiniBand CX6 200Gbps NIC  \\
\hline
Data SSDs           & 16 * 15.36TB PCIe 4.0x4 \\
\hline
\end{tabularx}
\label{storage_node}
\end{table}
\subsubsection{3FS Storage Node Hardware}
In Fire-Flyer 2 AI-HPC, we deployed 180 storage nodes, as shown in Table \ref{storage_node}, each node contains 16 PCIe 4.0 NVMe SSDs and 2 Mellanox CX6 200Gbps InfiniBand HCAs. With totally 360 * 200Gbps outbound InfiniBand HCAs, the system can total provide 9TB/s outbound bandwidh, and we actually achieved total \textbf{read throughput of 8TB/s}. The total 2880 NVMe SSDs provide over 20PiB storage space with an mirror data redundancy.

\subsubsection{Key Techinical Points of 3FS}\label{3fs_key_points}

The 3FS system comprises four roles: cluster manager, meta service, storage service and client. Meta and storage services send heartbeats to cluster manager. All services and clients poll cluster configuration and service status from the manager. Multiple cluster managers are present, with one elected as the primary.

File system meta data are stored in tables of a distributed key-value storage system. Each file or directory has a unique inode ID. The File inode/directory ID and meta data, such as file size and location information of the file content data, are stored as key-value pairs in the inode table. A separate directory entry table stores key-value pairs of $(parent\_dir\_inode\_id, entry\_name): (entry\_inode\_id, ...)$ to support iterating entries in a directory and resolving file/directory paths. All states of meta services are persisted on the distributed key-value storage system. Several meta services run concurrently to handle meta requests from clients.

The storage service has an implementation of Chain Replication with Apportioned Queries (CRAQ) \cite{craq_2009} to provide strong consistency. CRAQ's write-all-read-any approach helps to unleash the throughput and IOPS of all SSDs. File content are split into chunks, which are replicated over a chain of \textit{storage targets}. A \textit{chain table} contains an ordered set of chains. The meta service selects an offset in the chain table and a stripe size \textit{k} for each file. The file chunks are assigned to the next \textit{k} chains starting at the offset. To distribute read/write traffic evenly to all SSDs, each SSD serves multiple storage targets from different chains. The storage service runs on every storage node and manages a few storage targets.

The storage network has a Fat-Tree topology that provides full bisection bandwidth. By design, each 3FS client can access every storage service. At peak load, incast congestion is observed on the client side. \textbf{To mitigate this congestion, a request-to-send control mechanism is implemented} in storage service and client \cite{fds_2012}. After receiving a read request from a client, the service reads data from SSD and asks the client's permission to transfer the data. The client limits the number of concurrent senders. When a storage service is granted the permission to transfer, it sends the data with a RDMA WRITE followed by a RDMA SEND to notify the client. The request-to-send control increases end-to-end IO latency but it's required to achieve sustainable high throughput.

\subsubsection{3FS-KV}
3FS-KV is a shared-storage distributed data processing system built on top of 3FS, currently supporting three models: key-value, message queue, and object storage. It supports read-write separation and on-demand startup, allowing it to fully leverage the extremely high I/O throughput provided by 3FS. 3FS-KV supports DeepSeek's KV Context Caching on Disk technology \cite{context_caching}, which reduces the cost of LLM serving by an order of magnitude.

\subsection{HAI Platform: a Time-Sharing Scheduling Platform}

The principle of time-sharing scheduling is applied to cluster resource management. Users submit tasks, such as running Python / bash code, starting development containers, etc., and the platform interrupts and loads tasks according to current resource requirements, cluster busyness, etc. Task code needs to follow the platform coding rules to ensure that it can be continued from breakpoints, with the specific process as follows:

\begin{itemize}
  \item Accepting the interruption signal from the cluster;
  \item Saving checkpoints (model parameters, optimizer parameters, etc.);
  \item Notifying the cluster of interruptions;
  \item Recovering from the checkpoint and continuing to run.
\end{itemize}

The cluster deploying HAI Platform does not pool GPU resources, but classifies and marks them based on computing nodes as basic units, according to resource types, network areas, etc. The HAI Platform encourages users to fully utilize multiple GPUs simultaneously for parallel training, facilitating 99\% utilization.

\section{Stability and Robustness}\label{stability_section}

\subsection{Checkpoint Manager}
Training LLMs can span several months, during which unavoidable hardware failures may cause training interruptions. To minimize recovery time and support the HAI Platform's interrupt and recovery operations, we developed a checkpoint manager. Additionally, the substantial size of LLM checkpoints necessitated an efficient method for saving and loading them, leveraging the high throughput of 3FS. The checkpoint manager includes the following components:
\begin{itemize}
\item Parameters and optimization states are divided into chunks and written to 3FS using the 3FS batch write API, which is significantly faster than normal writes, achieving over 10 GiB/s per node. This enables saving to be completed in just a few seconds.
\item Parameters and optimization states are asynchronously transferred from GPU to CPU host memory, with checkpoint saving performed periodically (typically every 5 minutes).
\item During the saving process, each tensor is recorded with its index and the offset within the checkpoint., which makes the location of tensors more convenient during the loading process. With the 3FS batch read API, a loading process can be completed in just a few seconds.
\end{itemize}
Thanks to the high write throughput of 3FS, periodic saving operations can be completed asynchronously in a matter of seconds, without impacting the training process. In the event of hardware failures that interrupt training, only the last 5 minutes of progress are lost. For a cluster with thousands of nodes, this overhead from disaster recovery is minimal.

\subsection{Validator}
The best way to enhance device stability is to identify issues before they occur. Therefore, we have developed a set of validator tools to verify whether the hardware is functioning correctly. The platform's automatic operation and maintenance system runs the validator program weekly on nodes to verify their proper functionality. It removes the faulty nodes from the scheduling platform, ensuring that all scheduled nodes are operational. 
Diagnosing tools like hostping\cite{hostping_285185} also integrated in our platform, but to find root cause of Hardware Failures is still hard work for operation teams.
The validator mainly consists of the following parts:
\begin{itemize}
\item Checking hardware frequency, link speed, and link status.
\item Testing CPU stress and memory bandwidth.
\item GPU Memory test:  This involves checking each byte of GPU memory to ensure no data corruption has occurred.
\item Running GEMM with full GPU memory occupancy, which can simultaneously check whether there are any operational logic faults in the GPU chip.
\item Intra-node allreduce test: checking NVLink bandwidth through upper-level applications.
\item Storage bandwidth stress test to make sure storage is functioning normally.
\end{itemize}

\subsection{Hardware Failures Characterization in Fire-Flyer 2 AI-HPC}
In supercomputers and data centers, hardware failures and chip errors can lead to floating-point overflow, non-convergence, or slow convergence during model training\cite{he_understanding_2023}. This paper \cite{wang_understanding_2023} even directly points out that there is a substantial amount of Silent Data Corruption in data center processors, ultimately leading to a variety of complex issues that are difficult to replicate and locate. Indeed, in our practice, we have encountered computational errors and GPU memory errors not detected by Error Correction Code (ECC), which led to models' gradnorm spikes,  loss explosions and even non-convergence. How to tackle these hardware failures, promptly identify and categorize them, is a key issue to improve the online rate and overall utilization of cluster nodes.

\begin{table}[hbt]
\centering
\caption{Type of GPU Xid Errors and Its Causes}
\setlength{\tabcolsep}{5pt} 
\renewcommand{\arraystretch}{1.5} 

\newcolumntype{Y}{>{\raggedright\arraybackslash}p{0.1\linewidth}}
\newcolumntype{Z}{>{\raggedright\arraybackslash}X}

\begin{tabularx}{\linewidth}{|Y|Z|}
\hline
\multicolumn{1}{|p{0.2\linewidth}|}{Xid Errors}  & Analysis \\
\hline
\multicolumn{1}{|p{0.2\linewidth}|}{ Software Causes: Xid\_13/31 Xid\_43/45 } &  Triggered by application programs, software-related Xid messages  may indicate anomalies in GPU memory affecting code and data segments. However, it's crucial to consider other information for a comprehensive hardware functionality assessment. \\
\hline
\multicolumn{1}{|p{0.2\linewidth}|}{NVLink Error: Xid 74}  & Xid74 indicates errors in NVLink. For PCIe A100, it's mainly occurred on the NVLink Bridge between two GPUs. Its occurrence rate is several orders of magnitude higher than other hardware faults. Apart from stress testing to exclude those that are constantly repeating errors, there isn't a good way to avoid the occurrence of Xid74 issues. \\
\hline
\multicolumn{1}{|p{0.2\linewidth}|}{Memory ECC Error: Xid\_63/64 Xid\_94/95 } & Triggered when the GPU handles memory ECC errors on the GPU.  With the introduction of row remapping technology in A100, most instances can be resolved by simply resetting the GPU to retain optimal performance. \\
\hline
\multicolumn{1}{|p{0.2\linewidth}|}{Uncorrectable GPU Failures: Xid\_44/48 Xid\_61/62/69/79} & Thease failures mean an uncorrectable error occurs on the GPU, which is also reported back to the user application. A GPU reset or node reboot is needed to clear this error.  \\
\hline
\multicolumn{1}{|p{0.2\linewidth}|}{Other Failures: Xid 119} & Xid119 means GPU GSP module failed.
These failures need to do fieldiag test, and most need to
RMA.\\
\hline

\end{tabularx}
\label{type_of_xids}
\end{table}

\subsubsection{\textbf{GPU Xid Error}}\label{gpu_xid_section}
An Xid error\cite{nvidia_xid} is a general GPU fault message that originates from the NVIDIA driver, logged into the kernel or event log of the operating system. We have categorized various types of Xid errors and analyzed the potential causes that may lead to such errors, as shown in the Table \ref{type_of_xids}. 

Table \ref{all_xids_errors} in \nameref{Supplementary} shows the Xid errors that have occurred in our Fire-Flyer 2 AI-HPC over the past year. In our PCIe-based system, Xid74, also known as NVLink errors, account for a significant proportion, comprising 42.57\% of the total. This high frequency is due to the inherent fault rate of NVLink Bridge connectors, amplified by our extensive use of thousands of GPUs.

Software-related errors such as Xid13, Xid31, Xid43, and Xid45 suggest possible illegal memory access or instructions in user code. Notably, illegal memory access (Xid 43) accounts for 33.48\%, highlighting the need for improved memory management. However, these errors may also result from memory data corruption, so hardware faults should be considered if software bugs are ruled out.

\begin{figure}[tb]
    \centering
    \includegraphics[width=\linewidth]{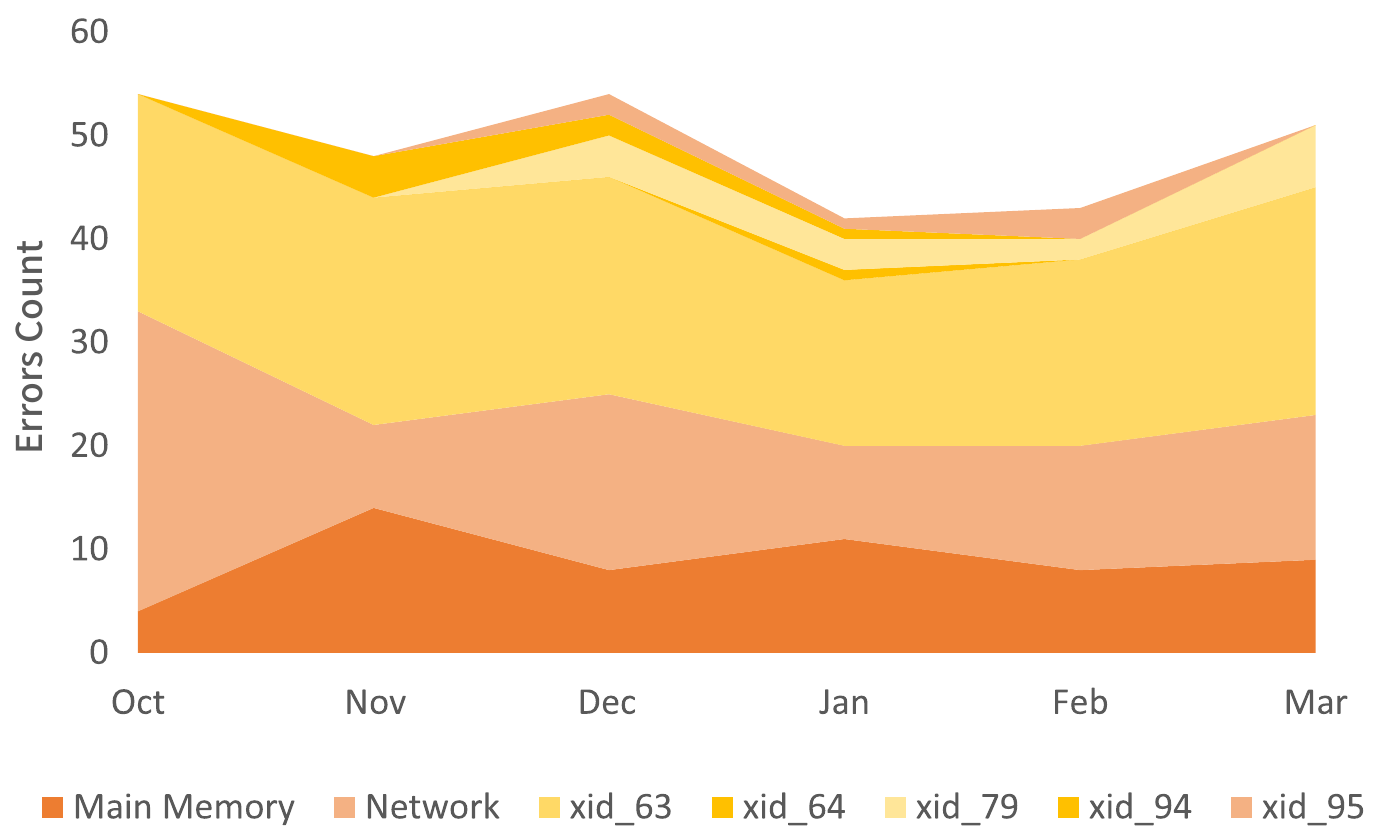}
    \caption{Trends of Memory and Network Failures from 2023 to 2024: ``Main Memory" indicates CPU Memory ECC errors, ``Network" indicates Network Flash Cuts, and ``xids" are related to GPU memory ECC errors. Raw data is available in  \nameref{Supplementary}, Table \ref{xid_and_network_raw_data}} .
    \label{xids_counts_trends}
\end{figure}

Additionally, GPU Memory ECC Errors,  such as Xid63, Xid64, Xid94, and Xid95, require special attention as they represent about 2\% of the total. Figure \ref{xids_counts_trends} illustrates the statistics related to ECC errors in our production cluster over the past six months. It is evident that the number of GPU ECC faults considerably surpasses those from the CPU. Therefore, it is crucial to promptly address GPU ECC faults to ensure that application performance and accuracy remain unaffected.

\subsubsection{\textbf{Network Flash Cut}}\label{ib_error_section}
\begin{figure}
    \centering
    \includegraphics[width=1\linewidth]{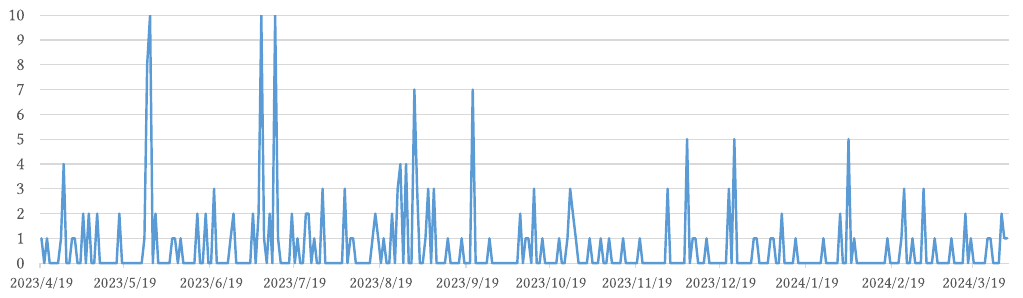}
    \caption{Trends of IB Network Failures: Link Flash Cuts}
    \label{ib_flagpping}
\vspace{-5pt} 
\end{figure}

In addition to CPU and GPU faults, network device malfunctions represent a significant portion of hardware issues. As shown in Figure\ref{xids_counts_trends}, IB link failures account for 30\% of hardware faults excluding Xid74. Network flash cuts can lead to application communication disruption, even task failures. Since most tasks run on multiple nodes, an issue on a single node can impact many others, further reducing cluster utilization. Figure \ref{ib_flagpping} illustrates the IB link failures data over the past year,  with raw data attached in  \nameref{Supplementary}, Table \ref{ib_raw_data},   indicating that these issues can occur \textbf{randomly} throughout the cluster's operational period.

\section{Discussion} \label{discuss_section}

\subsection{Discussion on Congestion Control in RDMA Networks}

Lossless RDMA networks offer several flow-control mechanisms, such as Priority Flow Control (PFC)\cite{2015PriorityFC} for RoCE networks and credit-based flow control\cite{doi:10.1142/S0219265906001843} for IB networks. In network routing, static routing algorithms in IB or ECMP (Equal-Cost Multi-Path)\cite{9377396} and AR (Adaptive Routing)\cite{9825915} effectively handle routing issues. However, congestion can still occur when multiple servers send data to a single receiver, potentially blocking the entire network. To mitigate this, IB NICs use DCQCN (Data Center Quantized Congestion Notification)\cite{10.1145/2829988.2787484} as their congestion control algorithm. While Data Processing Units (DPUs), such as the NVIDIA BF series, allow users to customize congestion control algorithms (like HPCC\cite{10.1145/3341302.3342085} and TIMELY RTT-based CC\cite{Mittal2015TIMELYRC}), they increase the cost and operational complexity of the cluster.

In practice, we chose to disable DCQCN to avoid its shortcomings, as it could not find parameters that simultaneously support HFReduce traffic and 3FS storage traffic in our Computation-Storage Integrated Network. Instead, we employed the network tuning methods mentioned in Section \ref{network_tuning_section}, ensuring our network operates without congestion control and remains congestion-free.

\subsection{Discussion about NVLink Technology Choices}
Initially, we did not use NVLink to avoid extra costs and maintain stability, as HFReduce was sufficient for training requirements at that time. However, as the demand for LLMs increased, we added NVLink specifically for LLM training purposes. The decision to install NVLink should be based on actual needs due to its potential drawbacks.

\subsection{Maintaince Cost Overview}

\subsubsection{Construction Cost}
Relative hardware costs are provided in Table \ref{pcie_cost} and \ref{cost_compare}. Software costs, contributed by several dozen in-house developers, are just a fraction of the cost for thousands of GPU servers.

\subsubsection{Power Consumption}
The average power consumption comparision during ResNet training is provided in Table \ref{pcie_cost}. Including the overhead from IB switches and other nodes, the total energy consumption of the Fire-Flyer 2 AI-HPC does not exceed 4 MW, approximately just over 3 MW.

\subsubsection{Operation Cost}
Operating costs can be estimated by considering power consumption and rack rental costs. By multiplying this figure by the number of nodes and the PUE (Power Usage Effectiveness), the total operating costs can be calculated.

\subsection{Stability Compared with Other Architectures}
A recent paper\cite{hu2024characterization} reportsthat NVLink-related failures account for approximately 52.42\% (54 out of 103) of total failures, with raw data indicating 54 NVLink Errors, 21 CUDA Errors, 16 Node Failures, 12 ECC Errors, and 12 Network Errors. In comparison, our NVLink-related issues, primarily Xid-74 Errors, as mentioned in Section \ref{gpu_xid_section}, account for about 42.57\% of GPU failures.

\section{Future Work}\label{nextgen_section}

\subsection*{Future Arch and Integration with New GPU Models}
\begin{figure}[bt]
    \centering
    \includegraphics[width=\linewidth]{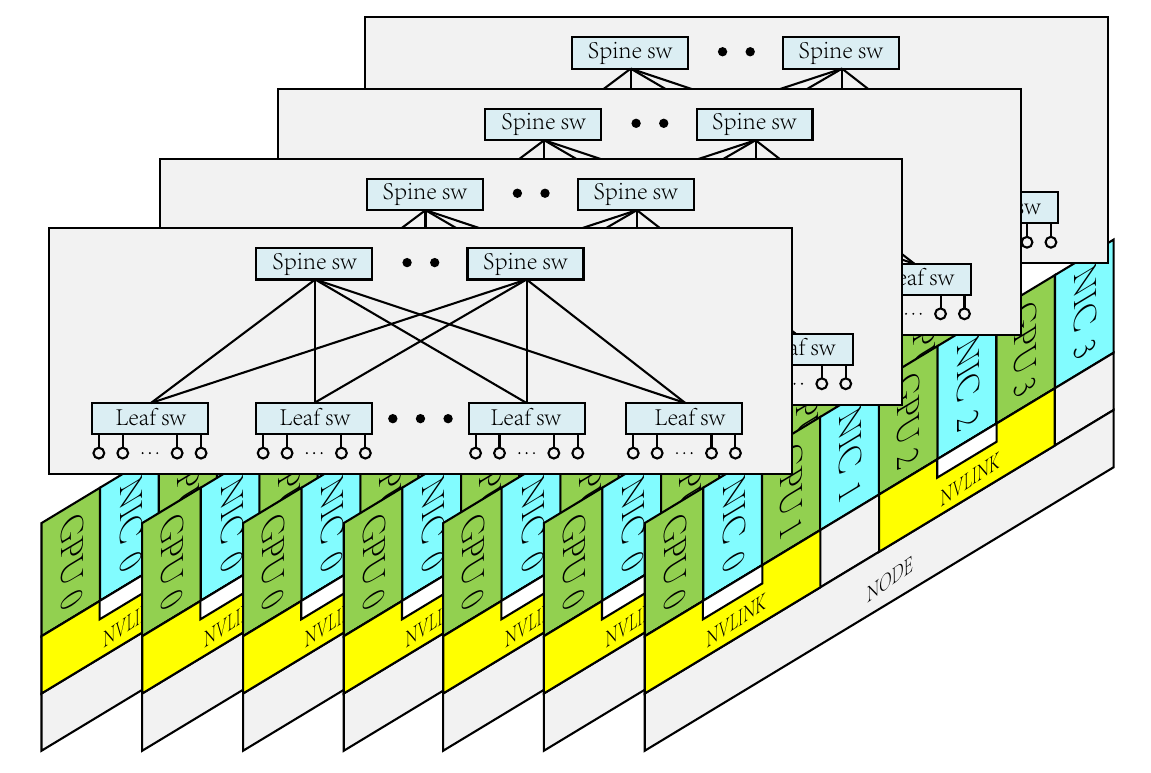}
\caption{Next Generation PCIe Node Atchitecture with Multi-Plane Fat-Trees Network}
\label{nextgen}
\vspace{-10pt} 
\end{figure}

 Our next-generation PCIe architecture is designed for MoE (Mixture of Experts) LLM training, where all-to-all performance is crucial. Therefore, the next-gen nodes feature a 1:1 GPU to NIC ratio, comparable to DGX-H100/B100 systems, as illustrated in Figure \ref{nextgen}.

We are considering implementing a multi-plane network to reduce costs while maintaining performance. Additionally, we are exploring the use of RoCE switches instead of IB switches, which can significantly lower network expenses. With a 128-port 400 Gbps RoCE switch, a 4-Plane Two-Layer Fat-Trees network can support up to 32,768 GPUs.

\section{Conclusions}
In this paper,  we have shared our experiences and insights from deploying and maintaining the Fire-Flyer 2 AI-HPC, which is equipped with 10,000 PCIe A100 GPUs.Our approach to PCIe architecture and storage-computation integrated network design has resulted in significant cost savings, effectively halving construction costs and demonstrating substantial cost-effectiveness.

In terms of software co-design, we introduced HFReduce and HaiScale to overcome hardware limitations, ensuring scalable performance of the PCIe architecture. Our in-house developed 3FS distributed file system,in conjunction with network co-design, facilitates traffic isolation for both 3FS and HFReduce allreduce traffic, effectively preventing congestion. The comprehensive software stack within the HAI Platform addresses a variety of system faults, from network congestion to hardware failures, thereby ensuring high stability and robustness.

Together, these software and hardware innovations enable our PCIe A100 architecture to achieve 80\% the performance of NVIDIA's DGX-A100, while consuming less than 60\% of its power. The practical knowledge we have accrued may prove valuable for both industrial and academic sectors. We hope that our work will serve as a reference for others aiming to build their own cost-effective and efficient AI-HPC clusters.

\section*{Acknowledgment}

We extend our heartfelt gratitude to all our colleagues at DeepSeek-AI and High-Flyer Quant for their invaluable contributions to the Fire-Flyer 2 AI-HPC project. Throughout the four years of design, construction, and operation, their collaborative efforts have been crucial in overcoming numerous challenges. Consequently, Fire-Flyer 2 AI-HPC now supports the training tasks of the DeepSeek-AI series large language models \cite{deepseek-ai_deepseek_2024,shao_deepseekmath_2024,lu_deepseek-vl_2024,deepseek-ai_deepseek-v2_2024,deepseek-ai_deepseek-coder-v2_2024,xin_deepseek-prover-v15_2024}. We extend a special thanks to the HFAiLab System Team and the Operations Team, whose core contributions have been essential to the success of this endeavor.

\bibliographystyle{IEEEtran}
\bibliography{conference_101719.bbl}

\clearpage

\appendices

\section*{Appendix: Supplementary Characterization} \label{Supplementary}

\begin{table}[ht]
\centering
\caption{\textbf{Raw Data} of GPU Xid Errors in Our Cluster Over the Past Year, as Mentioned in Section \ref{gpu_xid_section}}
\setlength{\tabcolsep}{5pt}
\renewcommand{\arraystretch}{1.5}

\newcolumntype{Y}{>{\raggedright\arraybackslash}p{0.1\linewidth}}
\newcolumntype{Z}{>{\raggedright\arraybackslash}X}

\begin{tabularx}{\linewidth}{|c|X|X|X|}
\hline
\textbf{GPU Error Type} & \textbf{Xid Code} & \textbf{Number} & \textbf{Percentage} \\
\hline
NVLink Error & xid\_74 & 5521 & 42.57\% \\
\hline
Software Causes & xid\_13 & 45 & 0.35\% \\
 & xid\_31 & 2487 & 19.18\% \\
 & xid\_43 & 4342 & 33.48\% \\
 & xid\_45 & 240 & 1.85\% \\
\hline
GPU ECC Error & xid\_63 & 245 & 1.89\% \\
 & xid\_64 & 2 & 0.02\% \\
 & xid\_94 & 13 & 0.10\% \\
 & xid\_95 & 17 & 0.13\% \\
\hline
Uncorrectable Failures & xid\_44 & 1 & 0.01\% \\
 & xid\_48 & 2 & 0.02\% \\
 & xid\_61 & 13 & 0.10\% \\
 & xid\_62 & 3 & 0.02\% \\
 & xid\_69 & 1 & 0.01\% \\
 & xid\_79 & 37 & 0.29\% \\
\hline
GPU GSP ERROR & xid\_119 & 1 & 0.01\% \\
\hline
\textbf{Total} & & \textbf{12970} & \textbf{100.00\%} \\
\hline
\end{tabularx}
\label{all_xids_errors}
\end{table}

\begin{table}[ht]
\centering
\onecolumn
\caption{\textbf{Raw Data} of Memory and Network Failures from 2023 to 2024, as Mentioned in section \ref{ib_error_section}, Figure \ref{xids_counts_trends}:\\ ``Main Memory" indicates CPU Memory ECC errors, ``Network" indicates Network Flash Cuts, and ``xids" are related to GPU memory ECC errors.}
\setlength{\tabcolsep}{5pt}
\renewcommand{\arraystretch}{1.6}

\newcolumntype{Y}{>{\raggedright\arraybackslash}p{0.1\linewidth}}
\newcolumntype{Z}{>{\raggedright\arraybackslash}X}

\begin{tabularx}{\linewidth}{|X|X|X|X|X|X|X|X|X|}
\hline
 & Main Memory & Network  & xid\_63 & xid\_64 & xid\_79 & xid\_94 & xid\_95 & Total \\
\hline
\multicolumn{9}{|c|}{2023} \\
\hline
October & 4 & 29 & 21 & 0 & 0 & 0 & 0 & 54 \\
November & 14 & 8 & 22 & 0 & 0 & 4 & 0 & 48 \\
Ddecember & 8 & 17 & 21 & 0 & 4 & 2 & 2 & 54 \\
\hline
\multicolumn{9}{|c|}{2024} \\
\hline
January & 11 & 9 & 16 & 1 & 3 & 1 & 1 & 42 \\
February & 8 & 12 & 18 & 0 & 2 & 0 & 3 & 43 \\
March & 9 & 14 & 22 & 0 & 6 & 0 & 0 & 51 \\
\hline
Total & 54 & 89 & 120 & 1 & 15 & 7 & 6 & 292 \\
\hline
\end{tabularx}
\label{xid_and_network_raw_data}
\end{table}

\begin{table}[t]
\caption{\textbf{Raw Data} of IB Network Failures and Flash Cuts Over the Past Year, as Mentioned in Section \ref{ib_error_section}, Figure \ref{ib_flagpping}. }
\setlength{\tabcolsep}{5pt} 
\renewcommand{\arraystretch}{1.5} 

\newcolumntype{Y}{>{\raggedright\arraybackslash}p{0.1\linewidth}}
\newcolumntype{Z}{>{\raggedright\arraybackslash}X}

\begin{tabularx}{\linewidth}{|X|X|X|X|X|X|X|X|}
        \hline
        Date & Failure Count & Date & Failure Count & Date & Failure Count & Date & Failure Count \\ \hline
        2023/4/19 & 1 & 2023/7/8 & 1 & 2023/9/7 & 3 & 2023/12/24 & 5 \\ \hline
        2023/4/21 & 1 & 2023/7/10 & 2 & 2023/9/12 & 1 & 2023/12/31 & 1 \\ \hline
        2023/4/26 & 1 & 2023/7/12 & 10 & 2023/9/17 & 1 & 2024/1/1 & 1 \\ \hline
        2023/4/27 & 4 & 2023/7/13 & 1 & 2023/9/21 & 7 & 2024/1/6 & 1 \\ \hline
        2023/4/30 & 1 & 2023/7/18 & 2 & 2023/9/27 & 1 & 2024/1/7 & 1 \\ \hline
        2023/5/1 & 1 & 2023/7/20 & 1 & 2023/10/8 & 2 & 2024/1/10 & 2 \\ \hline
        2023/5/4 & 2 & 2023/7/23 & 2 & 2023/10/10 & 1 & 2024/1/15 & 1 \\ \hline
        2023/5/6 & 2 & 2023/7/24 & 2 & 2023/10/11 & 1 & 2024/1/25 & 1 \\ \hline
        2023/5/9 & 2 & 2023/7/26 & 1 & 2023/10/16 & 1 & 2024/1/31 & 2 \\ \hline
        2023/5/17 & 2 & 2023/7/29 & 3 & 2023/10/22 & 1 & 2024/2/3 & 5 \\ \hline
        2023/5/26 & 1 & 2023/8/6 & 3 & 2023/10/25 & 1 & 2024/2/5 & 1 \\ \hline
        2023/5/27 & 8 & 2023/8/8 & 1 & 2023/10/26 & 3 & 2024/2/17 & 1 \\ \hline
        2023/5/28 & 10 & 2023/8/9 & 1 & 2023/10/27 & 2 & 2024/2/22 & 1 \\ \hline
        2023/5/30 & 2 & 2023/8/16 & 1 & 2023/10/28 & 1 & 2024/2/23 & 3 \\ \hline
        2023/6/5 & 1 & 2023/8/17 & 2 & 2023/11/2 & 1 & 2024/2/26 & 1 \\ \hline
        2023/6/6 & 1 & 2023/8/18 & 1 & 2023/11/6 & 1 & 2024/3/1 & 3 \\ \hline
        2023/6/8 & 1 & 2023/8/20 & 1 & 2023/11/9 & 1 & 2024/3/5 & 1 \\ \hline
        2023/6/14 & 2 & 2023/8/23 & 2 & 2023/11/14 & 1 & 2024/3/11 & 1 \\ \hline
        2023/6/16 & 0 & 2023/8/25 & 3 & 2023/11/20 & 1 & 2024/3/16 & 2 \\ \hline
        2023/6/17 & 2 & 2023/8/26 & 4 & 2023/11/30 & 3 & 2024/3/18 & 1 \\ \hline
        2023/6/20 & 3 & 2023/8/28 & 4 & 2023/12/7 & 5 & 2024/3/24 & 1 \\ \hline
        2023/6/26 & 1 & 2023/8/31 & 7 & 2023/12/9 & 1 & 2024/3/25 & 1 \\ \hline
        2023/6/27 & 2 & 2023/9/1 & 3 & 2023/12/10 & 1 & 2024/3/29 & 2 \\ \hline
        2023/7/4 & 2 & 2023/9/4 & 1 & 2023/12/14 & 1 & 2024/3/30 & 1 \\ \hline
        2023/7/6 & 2 & 2023/9/5 & 3 & 2023/12/22 & 3 & 2024/3/31 & 1 \\ \hline
        2023/7/7 & 10 & & & & & & \\ \hline
    \end{tabularx}
\label{ib_raw_data}

\end{table}

\end{document}